\documentclass[pra, 10pt, twocolumn, tightenlines, longbibliography, aps,
superscriptaddress,nofootinbib]{revtex4-2}

\usepackage[utf8]{inputenc}
\usepackage[T1]{fontenc}
\usepackage{float}
\usepackage[english]{babel}
\usepackage[normalem]{ulem}
\usepackage{mathtools}
\usepackage{bbold}
\usepackage{graphicx}
\usepackage{multirow}
\usepackage{tabularx}
\usepackage[dvipsnames, table]{xcolor}
\usepackage{soul}
\usepackage[colorlinks=true,allcolors=blue]{hyperref}
\usepackage{amsmath}

\DeclarePairedDelimiter\bra{\langle}{\rvert}
\DeclarePairedDelimiter\ket{\lvert}{\rangle}

\DeclarePairedDelimiterX\braket[2]{\langle}{\rangle}{#1 \delimsize\vert #2}

\begin{document}
	
\title{Phases and dynamics of an impurity immersed in one-dimensional quantum droplets}

\author{D. Diplaris}
\affiliation{Center for Optical Quantum Technologies, Department of Physics, University of Hamburg, Luruper Chaussee 149, D-22761 Hamburg, Germany}
\affiliation{The Hamburg Centre for Ultrafast Imaging, University of Hamburg, Luruper Chaussee 149, D-22761~Hamburg, Germany}

\author{I. A. Englezos}
\affiliation{Center for Optical Quantum Technologies, Department of Physics, University of Hamburg, Luruper Chaussee 149, D-22761 Hamburg, Germany}

\author{F. Theel}
\affiliation{Center for Optical Quantum Technologies, Department of Physics, University of Hamburg, Luruper Chaussee 149, D-22761 Hamburg, Germany}

\author{P. Schmelcher}
\affiliation{Center for Optical Quantum Technologies, Department of Physics, University of Hamburg, Luruper Chaussee 149, D-22761 Hamburg, Germany}
\affiliation{The Hamburg Centre for Ultrafast Imaging, University of Hamburg, Luruper Chaussee 149, D-22761~Hamburg, Germany}

\author{S. I. Mistakidis}
\affiliation{Department of Physics and LAMOR, Missouri University of Science and Technology, Rolla, MO 65409, USA}

\date{\today}

\begin{abstract} 

We explore the ground-state properties of a single impurity immersed in a one-dimensional quantum droplet medium formed by a two-component Bose mixture. 
Relying on ab initio simulations, we demonstrate that tuning the impurity--droplet interactions allows to controllably reshape the droplets' density profiles  and associated correlation patterns. 
For attractive impurity-medium couplings, the impurity becomes localized within the droplet, which exhibits a density hump at the vicinity of the impurity, while repulsive interactions {facilitate phase separation}. 
Comparing our many-body results {with} the appropriate extended Gross–Pitaevskii description, we find adequate agreement for the droplet density profiles, with the effective field approach systematically overestimating impurity localization. 
Following a release of the external trap, we unveil that the sign and magnitude of the interactions between the impurity and the droplet hosts dictate the response of the three-component setting, which experiences expansion unless strongly attractive intercomponent couplings are present. 
These results corroborate the role and presence of correlations in impurity--droplet mixtures and inspire future investigations on impurity physics for probing droplet configurations.

\end{abstract}

\maketitle

\section{Introduction}

Quantum droplets represent an emergent class of liquid type many-body self-bound states of matter~\cite{luo2021new,bottcher2020new,bose-gases-low-d-mistakidis}. 
They are stabilized due to the delicate interplay between attractive (repulsive) mean-field interactions and repulsive (attractive) quantum fluctuations in three- (one-) dimensions~\cite{petrov-droplets-2015-PhysRevLett.115.155302,petrov-ultra-dilute-2016-PhysRevLett.117.100401}. {On the contrary, in two-dimensions the mean-field coupling and quantum fluctuation contribution are incorporated in a logarithmic nonlinear term associated with attractive (repulsive) effective interactions at low (large) densities, thereby giving rise to a different stabilization mechanism~\cite{petrov-ultra-dilute-2016-PhysRevLett.117.100401}.} 
These states were first observed in anisotropic dipolar gases~\cite{ferrier2016observation,chomaz2022dipolar} and subsequently in short-range interacting bosonic mixtures~\cite{CabreraTarruellDropExp,CheineyTarruellDropExp,SemeghiniFattoriDropExp,GuoHeteroDrop2021,FortHeteroExp,Cavicchioli}.  
Specifically, quantum corrections are often modeled to first-order beyond the mean-field approximation by the Lee-Huang-Yang (LHY) energy term~\cite{LeeHuangYang1957}. The latter depends on the dimension~\cite{dim_crossover_Zin,Ilg_crossover_2018,Pelayo_crossover} and the effective range of the involved interactions~\cite{zhang2025self}, yielding associated extended Gross-Pitaevskii equations (eGPEs)~\cite{petrov-droplets-2015-PhysRevLett.115.155302,petrov-ultra-dilute-2016-PhysRevLett.117.100401,ilias-simos-SciPostPhys.19.5.133}. 

This framework has been extensively used to capture a distinctive landscape of droplet properties. 
In short-range two-component bosonic mixtures, these include flat-top density profiles~\cite{1Ddrops_stat_dyn}, mixed droplet phases~\cite{ilias-simos-particle-imbalance-PhysRevA.110.023324,Kartashov_multipoles,xiao2026one} with a gaseous fraction~\cite{flynn2023quantum,flynn2024harmonically,Pelayo_2025}, and involved  excitation spectra~\cite{spectrum1D,Katsimiga_sol_drops,charalampidis2024two}. 
Notably, enriched nonlinear dynamics hosting vortical patterns~\cite{Li_vortex,Yoifmmode,Tengstrand_vortex,Bougas_vortex},  shock~\cite{Chandramouli_shocks} and rogue~\cite{Chandramouli_RWs} waves due to the modulationally unstable character~\cite{mithun2020modulational,Otajonov_MI_drops}, as well as kink configurations~\cite{kartashov2022spinor,katsimiga2023interactions} that are also stable in two-dimensions~\cite{Mistakidis_kink} have been revealed. 
Additionally, the role of beyond-LHY correlations has been investigated in two-component mixtures using either ab initio approaches~\mbox{\cite{ParisiGiorginiMonteCarlo,ParisiMonteCarlo2019,droplet-harmonic_pot-Mistakidis_2021,ilias-simos-droplet-excitations-PhysRevA.107.023320}} or perturbative techniques~\cite{Hu_theory,Gu_phonon,Pan_critical,Ota_beyond_LHY}, revealing, for instance, deviations from the eGPE predictions at mesoscopic atom numbers and stronger attractions.

On the other hand, multicomponent quantum droplets open another frontier for engineering unseen self-bound many-body phases due to their arguably enlarged parameter space delimited by the introduction of additional interaction strengths, atom numbers, and masses~\cite{ilias-simos-SciPostPhys.19.5.133,englezos2026stabilitymixedphasesthreecomponent}. 
Here, prototypical settings constitute three-component attractively interacting mixtures. 
Their properties are much less explored~\cite{ma2021borromean,ilias-simos-SciPostPhys.19.5.133,englezos2026stabilitymixedphasesthreecomponent,Cui2025_3CompDroplets,Edler_buoyancy} and have been very recently shown to accommodate, especially for particle imbalanced systems, exotic mixed droplet phases~\cite{englezos2026stabilitymixedphasesthreecomponent} that are not present in their two-component counterparts. 
Particularly, in the limit of extreme particle imbalance such systems can be utilized to controllably study impurity physics (captured by one of the components) within a quantum droplet represented by the other two components.  
This concept has been extensively studied in the past within the realm of repulsive Bose gases~\cite{scazza2022repulsive,bose-gases-low-d-mistakidis,Grusdt_2025} and argued to be inherently related to quantum fluctuations facilitating the polaron generation.

In this vein, the immersion of an impurity into a quantum droplet is far less understood, while the impact of beyond-LHY effects and ensuing correlation patterns, which is the focus of our study, remains an open question. 
Recent investigations unveiled mixed droplet-impurity excitation modes~\cite{sinha-imp-1d-droplet}, self-localization of either the impurity~\cite{abdullaev2020bosonic,Bighin_localization} or the droplet due to a heavy impurity~\cite{debnath2023interaction,Bristy_defect_drops}, tunability of the dipolar bound state character~\cite{wenzel2018fermionic}, and mediated interactions of fermionic impurities from the bosonic droplet~\cite{Pelayo_ferm_imp}. 
It is, however, yet unclear how an impurity can regulate the two-component droplet as well as which phases can be formed in this setting and whether they are entirely self-bound. 
To explore these possibilities here we employ a three-component mixture consisting of an impurity embedded into a genuine two-component droplet in one-dimension (1D). 
Our investigation relies on the suitable, recently constructed eGPEs~\cite{ilias-simos-SciPostPhys.19.5.133} and the ab initio multi-layer multi-configuration time-dependent Hartree method for atomic mixtures (ML-MCTDHX)~\cite{cao2017}, which enables us to capture beyond-LHY correlations. 

We find that the impurity acts as a knob for reshaping the droplet density profile and associated correlation configurations. 
Focusing on symmetric impurity--droplet couplings (i.e., both repulsive or attractive), it is shown that for attractive interactions, the impurity features spatial localization. 
This is  accompanied by a prominent density hump at the droplet core since atoms from the droplet accumulate in the vicinity of the impurity. 
Note that the deformation of the droplet profile is substantially different {from} the case of an attractive potential well {that mimics} an infinitely heavy impurity where an analytical droplet solution can be extracted~\cite{debnath2023interaction,Bristy_defect_drops}. 
However, repulsive impurity--droplet interactions facilitate impurity expulsion and a transition to phase-separation between the impurity and the droplet. 
A similar back-action of the impurity to the density of the majority component has been reported for repulsive Bose gases~\cite{bose-gases-low-d-mistakidis,Grusdt_2025} but it is yet to be explored for droplet hosts. 
Turning to mixed impurity--droplet couplings, we explicate that they allow to selectively deform the density of the individual droplet components. 
They yield more complex three-component configurations characterized by spatially localized humps and dips in the droplet-components for repulsive and attractive couplings, respectively. 

Direct comparisons between the eGPE predictions and the ab initio simulations reveal that the former qualitatively reproduce the deformed and phase-separated impurity and droplet states. 
However, they overestimate the degree of impurity localization and predict {an} earlier onset of phase separation, suggesting that beyond-LHY correlations enhance the droplet's ability to accommodate repulsive impurities. 
Additionally, the associated two-body coherences{, which naturally can not be estimated with the eGPE model,} reveal that the characteristic anti-bunching behavior of the droplet~\cite{ilias-simos-droplet-excitations-PhysRevA.107.023320,ilias-simos-particle-imbalance-PhysRevA.110.023324} is preserved for weak impurity--droplet couplings, but becomes modified {at stronger interaction} strengths, giving rise to more complicated correlation patterns. 
Finally, we explore the droplet dynamics in the mixed impurity--droplet coupling regime following a trap release in all components. 
It turns out that the emergent dynamical response is determined by both the sign and strength of the impurity-host interactions. 
{Specifically}, the majority components featuring strong attraction with the impurity maintain their shape, otherwise they undergo expansion.

This work is structured as follows. 
Section~\ref{sec:theory} introduces the three-component bosonic mixture, consisting of a two-component droplet and an impurity, while Section~\ref{sec:approaches} elaborates on the ab initio approach and the appropriate eGPEs used to study the system under consideration. 
In Section~\ref{sec:ground_conf}, we analyze the ensuing ground-state configurations when the impurity couples with the individual droplet components with the same or opposite interaction signs. 
Section~\ref{sec:dynamics} demonstrates the expansion dynamics of selective droplet--impurity structures within the eGPE framework after being released from their trap. 
Section~\ref{sec:conclusions} summarizes the important findings of our work and discusses future perspectives emanating from our results. 
Appendix~\ref{app:impurity:freespace} explicates the impact of the impurity on the majority component in a box potential, while Appendix~\ref{app:dressing} elaborates on the impurity dressing by its hosts through a suitable fidelity measure.


\section{Impurity Embedded in a Two-Component Droplet}\label{sec:theory}

We consider a homonuclear ($m_A=m_B=m_C \equiv m$) 1D three-component bosonic mixture in the ultracold regime. 
Accordingly, interatomic interactions can be modeled by contact interaction potentials since s-wave scattering processes dominate~\cite{Pethick_Smith_2008}. 
Additionally, we assume that species A and B contain $N_A=N_B=20$ atoms which act as a bath for the single impurity ($N_C=1$) represented by species C. 
The underlying many-body Hamiltonian of this system reads: 
\begin{align} \label{MB_Hamilt}
\begin{split}
H = \sum_{\sigma} H_{\sigma} + \sum_{\sigma \neq \sigma^\prime} H_{\sigma \sigma^\prime},
\end{split}
\end{align}
where $\sigma=A,B,C$. 
The intra- and interspecies Hamiltonians of Equation~(\ref{MB_Hamilt}) are given by:
\begin{subequations}
\begin{align}
H_{\sigma} &= \sum_{i=1}^{N_{\sigma}} h^{(1)}_{\sigma} (x_i) + g_{\sigma \sigma} \sum_{i<j} \delta (x_i^{\sigma} - x_j^{\sigma}), \\
H_{\sigma \sigma^\prime} &= g_{\sigma \sigma^\prime} \sum_{i=1}^{N_\sigma} \sum_{j=1}^{N_{\sigma^\prime}} \delta (x^{\sigma}_i - x^{\sigma^\prime}_j ), \quad \sigma \neq \sigma^{\prime}
\end{align}
\end{subequations}
respectively, where $h^{(1)}_{\sigma}(x_i) = -\frac{\hbar^2}{2m} \frac{\partial^2}{\partial x^2} + \frac{1}{2} m \omega^2 x_i^2$ stands for the one-body Hamiltonian of each species. 

The above 1D setup can be experimentally realized by employing a sufficiently strong confinement along the perpendicular directions $y, z$. 
This translates to the condition $\sqrt{\omega_y\omega_z}=\omega_{\perp}\gg \omega$, allowing to kinematically constrain the atoms along the elongated $x$ direction, which is customarily done in  corresponding 1D experiments~\cite{Ketterle2001LowDexp,Romero_exp_Peregrine}. 
Moreover, the coupling strengths can be experimentally tuned via magnetic~\cite{feshbach1-RevModPhys.78.1311, feshbach2-RevModPhys.82.1225}, or confinement induced~\cite{confinement-resonance-1-PhysRevLett.81.938} Fano--Feshbach resonances.

For computational convenience, we rescale the Hamiltonian with respect to $\hbar \omega_{\perp}$. 
As such, the time, length and interaction strengths are measured in units of $\omega_{\perp}^{-1}$, $\alpha_{\perp}=\sqrt{\hbar / (m \omega_{\perp})}$ and $\sqrt{\hbar^3\omega_{\perp}/m}$, respectively. 
Also, hard-wall boundary conditions are imposed at $x= \pm 90$, which is large enough to prevent finite size effects. 
Our system is harmonically trapped featuring $\omega=0.005$, which is sufficiently weak to enable comparisons with the suitable LHY theory framework (Section~\ref{sec:LHY}) that is valid within the local density approximation~\cite{petrov-ultra-dilute-2016-PhysRevLett.117.100401,ilias-simos-SciPostPhys.19.5.133}.  
Typical length and time scales for the droplets to be examined in what follows are in the dimensionless range of $x \in [-50, 50]$ and $t \in [0,10^3]$, which refer to $x \in [-24,24]$ $\mu$m 
 and $t \in [0, 382]~{\rm ms}$, respectively, for a transverse confinement of $\omega_{\perp}=2 \pi \times 500~{\rm Hz}$.

Since we are interested {in studying} the physics of an impurity {embedded in droplet hosts,} we fix $g_{AA}=g_{BB}=0.1$ and $g_{AB}=-0.02$, satisfying $\delta g=g_{AB}-\sqrt{g_{AA}g_{BB}}>0$. This ensures that species A, B enter the 1D droplet regime~\cite{petrov-ultra-dilute-2016-PhysRevLett.117.100401,bose-gases-low-d-mistakidis} and in particular form flat-top droplet profiles (see also Section~\ref{sec:Decoupled}). 
Furthermore, the couplings $g_{AC}$, $g_{BC}$ between each bath component and the impurity are varied to explore the impact of the impurity on the droplet hosts. 
Particularly, below we consider both the cases of symmetric (i.e., $g_{AC}=g_{BC}$) and mixed (namely $g_{AC} \neq g_{BC}$) impurity--droplet couplings in Section~\ref{sec:Symmetrical} and Section~\ref{sec:Mixed}, respectively.  
In the former scenario, the A and B components form a so-called symmetric droplet~\cite{petrov-ultra-dilute-2016-PhysRevLett.117.100401,bose-gases-low-d-mistakidis} due to the specific choice of atom number and interaction parameters, while in the latter setup, the aforementioned symmetry breaks and components A, B behave differently,  assembling in a genuine two-component droplet~\cite{charalampidis2024two}. 
In a corresponding forthcoming experiment, this three-component system can be, for instance, realized by three distinct hyperfine states of $^{39}$K, two of which
have already been exploited in three-dimensional droplet settings~\cite{SemeghiniFattoriDropExp,CabreraTarruellDropExp}. 
Interestingly, 1D droplet configurations are yet to be experimentally observed in ultracold quantum simulators, an aspect that further motivates our investigation.



\section{Many-Body Theoretical Approaches}\label{sec:approaches}

It is well established that the formation of droplet configurations, taking place at attractive intercomponent interactions, requires the presence of quantum corrections beyond the mean-field approximation~\cite{luo2021new,bose-gases-low-d-mistakidis}. 
Below, we briefly summarize the many-body approaches 
{that operate} at different correlation levels and {are} utilized in our work to explore the role of the impurity in droplet hosts.

\subsection{Variational Method}\label{sec:ML-X}

To address the correlated many-body ground-state of the three-component system described by  Equation~\eqref{MB_Hamilt}, we employ the ab initio 
 ML-MCTDHX method~\cite{cao2017, kronke2013}.
A central facet of this approach is the expansion of the full many-body wave function in a multi-layer structure exploiting time-dependent and variationally optimized basis functions. 
This expansion facilitates the optimal truncation of the ensuing Hilbert space while accounting for
the relevant intra- and intercomponent correlations of cold atom settings (see also the Reviews~\cite{bose-gases-low-d-mistakidis, mctdh-colloquium-RevModPhys.92.011001} for elaborated discussions and applications in disparate systems).  
ML-MCTDHX allows us to tackle multicomponent bosonic  ultracold settings, but for clarity, we restrict ourselves to three-component Bose mixtures relevant for our study.  

First, at the top layer, the full quantum many-body wave function is expressed in a truncated basis of $D_{\sigma}$, orthonormal time-dependent species functions, $\ket{\Psi^{\sigma}_i(t)}$. 
This expansion enables us to account for intercomponent correlations and reads:
\begin{align}
| \Psi^{MB} (t) \rangle = \sum_i^{D_A} \sum_j^{D_B} \sum_k^{D_C} A_{ijk}(t) |\Psi_i^A (t) \rangle |\Psi_j^B (t) \rangle |\Psi_k^C (t) \rangle, 
\label{eq:psi_mlx}
\end{align}
where $A_{ijk}(t)$ are the time-dependent expansion coefficients. 
They correspond to the eigenvalues of the $N_{\sigma}$ species reduced density matrices and provide information about different intercomponent entanglement processes~\cite{vidal2002} through suitable entropy measures (see also Ref.~\cite{Theel_crossover} for corresponding discussions).

{Next, intracomponent correlations are incorporated in the second layer.}
To do so, each of the species functions $\ket{\Psi_i^{\sigma}(t)}$ is  expanded with respect to time-dependent bosonic number states $\ket{\vec{n}^{\sigma}_t}$, weighted by time-dependent coefficients $C_{i,\vec{n}^{\sigma}}^{\sigma} (t)$, yielding:
\begin{align}
|\Psi^{\sigma}_i (t) \rangle = \sum_{\vec{n}|N_{\sigma}}  C_{i,\vec{n}^{\sigma}}^{\sigma} (t) |\vec{n}^{\sigma} (t)\rangle. 
\end{align}
In this expression, the summation runs over all ${N_{\sigma} + d_{\sigma} -1}\choose{d_{\sigma} - 1}$ possible number state configurations with the $N_{\sigma}$ bosons distributed in $d_{\sigma}$ time-dependent single-particle functions (SPFs) $\ket{\phi_j^{\sigma}(t)}$. 
The vector $\vec{n}^{\sigma} = (n_1^{\sigma}, \dots, n_{d_{\sigma}}^{\sigma})$ designates the occupation number of each SPF. 
Finally, in the third layer, the SPFs are expanded on a time-independent basis comprising of $\mathcal{M}$ grid points, using  the discrete variable representation.  
Subsequently, the ML-MCTDHX equations of motion for the above-described coefficients are derived, e.g., by using the Dirac--Frenkel~\cite{frenkel1934wave} variational principle, determined by $\langle \delta \Psi | (i \hbar \partial_t - \hat{H} ) | \Psi \rangle = 0$. 
In order to compute the ground-state of the three-component mixture, the resulting equations are numerically solved using the standard imaginary-time propagation method. 
For our implementation, we deploy $\mathcal{M} = 1500$ grid points within the interval $[-90, 90]$, $(N_A,N_B,N_C)=(20,20,1)$ bosons, while the number of species functions and SPFs correspond to $(d_A,d_B,d_C)=(8,8,8)$,  and $(D_A,D_B,D_C)=(4,4,8)$, respectively.
This orbital configuration space ensures the numerical convergence of our results. 
{Let us also note that the expansion of the system’s many-body wavefunction, within ML-MCTDHX, in terms of a time-dependent and variationally optimized basis allows to efficiently simulate different number of atoms or interactions or masses per component, while utilizing a
computationally feasible basis size. 
This means that the number of the resulting equations of motion that need to be solved should remain numerically tractable and simultaneously numerical convergence should be ensured. 
Discussions about a wide range of applications of the method in different systems can be found in the Reviews~\cite{bose-gases-low-d-mistakidis,mctdh-colloquium-RevModPhys.92.011001}.}

We remark that for a given number of grid points $\mathcal{M}$, ML-MCTDHX becomes numerically exact when $d_{\sigma}=\mathcal{M}$ and $D_{\sigma}=\binom{N_{\sigma} + d_{\sigma} - 1}{d_{\sigma} - 1}$~\cite{cao2017}. 
In the opposite limit of $D_{\sigma}=d_{\sigma}=1$, the ML-MCTDHX wave function ansatz  reduces to the standard mean-field product ansatz~\cite{Pethick_Smith_2008} and the variational principle leads to the respective coupled  Gross-Pitaevskii equations of motion (see also Equation~(\ref{3comp_eGPEs}) with $\mathcal{E}_{LHY}=0$ and the discussion in Section~\ref{sec:LHY}). 


\subsection{Extended Gross-Pitaevskii Equations for the Impurity--Droplet System}\label{sec:LHY}

The impact of quantum fluctuations in bosonic mixtures can also be approximately captured using perturbation theory~\cite{petrov-droplets-2015-PhysRevLett.115.155302,petrov-ultra-dilute-2016-PhysRevLett.117.100401,ilias-simos-SciPostPhys.19.5.133}.
In particular, the first-order quantum correction to the mean-field energy functional, known as the LHY energy~\cite{petrov-droplets-2015-PhysRevLett.115.155302}, has been recently shown to adequately describe droplet formation. 
For an untrapped homonuclear two-component mixture in 1D, the LHY energy has been calculated {by employing a Bogoliubov ansatz}~\cite{petrov-ultra-dilute-2016-PhysRevLett.117.100401,ilias-simos-SciPostPhys.19.5.133} and across the mean-field stability regime, it is given by: 
\begin{equation}\label{E-LHY}
\mathcal{E}_{LHY} = \frac{\sqrt{m}(g_{AA} n_A + g_{BB} n_B)^{\frac{3}{2}}}{2\hbar \pi} \mathcal{I}(p),
\end{equation}
with $p=\frac{4(g_{AB}^2-g_{AA}g_{BB}) n_A n_B}{(g_{AA} n_A + g_{BB} n_B)^2}$, and 
\begin{equation}\label{I(p)}
    \mathcal{I}(p) = - \frac{\sqrt{2}}{3} 
    \Big( 
    (1 - \sqrt{p+1})^{\frac{3}{2}} + (1 + \sqrt{p+1})^{\frac{3}{2}}
    \Big).
\end{equation}
For completeness, we highlight that in the 1D system considered herein, the LHY correction term, $\mathcal{E}_{LHY}$, is attractive and has a different form from the one in higher dimensions~\cite{petrov-droplets-2015-PhysRevLett.115.155302} as well as long-range interacting settings~\cite{Chomaz_QFD,chomaz2022dipolar}.

To incorporate the bosonic impurity immersed in the two-component droplet host, we consider the standard mean-field coupling between the impurity and the droplet components described by $g_{\sigma C}\lvert \Psi_C\rvert^2 $~\cite{Pethick_Smith_2008,pitaevskii2003bose}.
Also, as mentioned above, we assume a sufficiently weak harmonic confinement characterized by $\omega=0.005$. 
This ensures that the LHY energy of Equation~(\ref{E-LHY}), calculated within the local density approximation, remains approximately valid~\cite{petrov-ultra-dilute-2016-PhysRevLett.117.100401,ilias-simos-SciPostPhys.19.5.133}.
Accordingly, following a variational principle, it is possible to extract the underlying coupled system of eGPEs~\cite{ilias-simos-SciPostPhys.19.5.133,englezos2026stabilitymixedphasesthreecomponent}

{\small
\begin{subequations}\label{3comp_eGPEs}
\begin{align}
i\hbar\frac{\partial \Psi_\sigma}{\partial t} &= \Bigg( h_0+ g_{\sigma} \lvert \Psi_{\sigma} \rvert ^2 + g_{AB} \lvert \Psi_{\sigma^\prime \neq \sigma} \rvert ^2 +g_{\sigma C}\lvert \Psi_C\rvert^2 \notag\\
&\quad\quad\quad  + \frac{\partial \mathcal{E}_{LHY}}{\partial n_{\sigma}} \Bigg) \Psi_{\sigma}, \label{eGPE_major} \\ 
i\hbar \frac{\partial \Psi_C}{\partial t} &= \Bigg( h_0 + g_{AC} |\Psi_{A}|^2 + g_{BC} |\Psi_{B}|^2 \Bigg) \Psi_C, 
\end{align}
\end{subequations}
}
where $h_0= - \frac{\hbar^2}{2m} \frac{\partial^2}{\partial x^2} + \frac{m\omega^2}{2}x^2$ and the species index $\sigma, \sigma^\prime = A, B$. 

Note that in the absence of the impurity species, the above expressions are valid for any interaction strength and particle number within the mean-field stability regime{, i.e., as long as $g_{AA}g_{BB}\geq g_{AB}^2$}~\cite{ilias-simos-SciPostPhys.19.5.133}. 
In the following, we will compare the predictions of this eGPE model with the exact many-body results obtained with the ML-MCTDHX approach in order to explore the role of the impurity in droplet hosts. 
By doing so, we are also aiming to benchmark the accuracy of the eGPEs at the mesoscopic particle number {regime} and identify the impact of beyond-LHY correlations. 
{Recently, it has been showcased that the eGPE predictions for a two-component droplet containing mesoscopic particle numbers in the absence of the impurity~\cite{ParisiMonteCarlo2019,droplet-harmonic_pot-Mistakidis_2021} are adequate when compared to corresponding many-body simulations.}
{Let as also note explicitly that the Bogoliubov ansatz, which is the basis for the derivation of the LHY energy, does not include  inter-component correlations, i.e., it corresponds to the case $D_A=D_B=D_C=1$ in Equation~\eqref{eq:psi_mlx}. 
Moreover, within the LHY treatment, only the intra-component correlations are taken into account up to first order, and hence higher order contributions are absent in the eGPE model. 
In addition, the LHY energy is calculated in free space and as such the impact of the external geometry to the correlations is neglected, while it is naturally captured within the ML-MCTDHX approach.}
Finally, it can be readily seen that if the LHY correction is neglected the above eGPE model reduces to the standard coupled set of three mean-field equations of motion~\cite{Pethick_Smith_2008}.


\begin{figure}
\centering
\includegraphics[width=\linewidth]{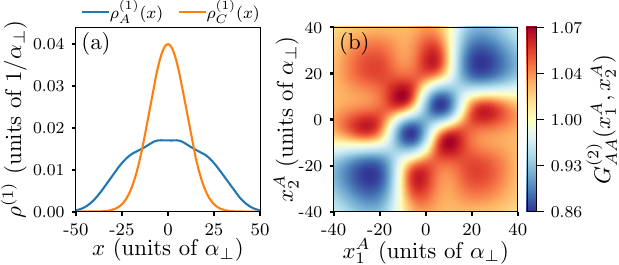}
\caption{(a) Ground{$-$}state density distributions of a symmetric two{$-$}component 
bosonic droplet and an impurity (see legends), at the decoupled limit (i.e., $g_{AC}=g_{BC}=0$). The droplet density profile, $\rho^{(1)}_A(x)=\rho^{(1)}_B(x)$, exhibits a flat{$-$}top spatial configuration, while the impurity, $\rho^{(1)}_C(x)$, has a Gaussian profile localized around the trap center, $x=0$.
(b) Intraspecies two{$-$}body coherence function, $G^{(2)}_{AA}(x_1^A, x_2^A)$, of the droplet component $A$. 
The characteristic droplet correlation pattern characterized by anti{$-$}bunching (bunching) across the diagonal (off{$-$}diagonal)
occurs for the droplet (namely majority) components.  
The remaining system parameters refer to $g_{AA}=g_{BB}=0.1$, $N_A=N_B=20$, $g_{AB}=-0.02$ and $\omega=0.005$.  
}
\label{fig:fig1-decoupled-impurity}
\end{figure}

\section{Ground State Phases}\label{sec:ground_conf}

In the following, we examine the ground-state properties of the three-component harmonically trapped ($\omega=0.005$) bosonic mixture for different droplet--impurity coupling strengths ($g_{AC}$, $g_{BC}$). 
Throughout, $g_{AA}=g_{BB}=0.1$ and $g_{AB}=-0.02$ are held fixed, favoring the flat-top droplet distribution for components $A$ and $B$ which accommodate $N_A=N_B=20$ bosons.  

We begin (Section~\ref{sec:Decoupled}) by briefly addressing the setting of a decoupled impurity, i.e., $g_{AC}=g_{BC}=0$, which reduces to a separable system of a two-component droplet in components $A$, $B$ and a single impurity in component $C$. 
Afterwards, in Section~\ref{sec:Symmetrical}, the case of a species $C$ impurity symmetrically coupled (namely $g_{AC}=g_{BC}$) with the two-component droplet is discussed. 
Next, in Section~\ref{sec:Mixed}, we elaborate on the most general system characterized by mixed impurity--droplet coupling strengths, i.e., $g_{AC}\neq g_{BC}$. 
In all cases, we primarily inspect the density profiles of the resulting structures building atop the individual components and their associated correlation patterns.  
Our analysis is mainly based on the ML-MCTDHX approach, which we also compare to the LHY approximation predictions where appropriate.


\subsection{Decoupled impurity-droplet setting}\label{sec:Decoupled}

We start by considering the limit of a $C$ impurity decoupled from the $A$, $B$ droplet components, meaning that $g_{AC}=g_{BC}=0$.
As such, the majority components are expected to arrange into a droplet configuration due to the chosen interaction strengths satisfying $\delta g=g_{AB}+\sqrt{g_{AA}g_{BB}}>0$, and in fact behave identically because of the equal atom numbers, masses, and intracomponent interactions~\cite{petrov-ultra-dilute-2016-PhysRevLett.117.100401}. 
This setting serves as a baseline for our main investigations with finite impurity--droplet couplings to be presented below. 

To visualize the ground-state spatial distribution of each component, we resort to the diagonal of the $\sigma$-species one-body reduced density matrix~\cite{rdm-sankmann-2008-PhysRevA.78.023615,correl-effects-2species-Mistakidis_2018} referring to the respective one-body densities 
\begin{align}
\rho^{(1)}_{\sigma} = \langle \Psi^{\rm MB} | 
\hat\Psi^{\dagger}_{\sigma} (x) \hat\Psi_{\sigma} (x)
| \Psi^{\rm MB}  \rangle.
\end{align}
{Here} 
, $\hat\Psi^{\dagger}_{\sigma} (x)$ ($\hat\Psi_{\sigma} (x)$) denotes the bosonic field operator that creates (annihilates) a boson of species $\sigma$ at position $x$. 
Figure~\ref{fig:fig1-decoupled-impurity}a illustrates the ground-state one-body density for the majority component $A$ (recall that component $B$ is the same and not shown for brevity) and the impurity $C$. 
As expected, the majority components assemble in a flat-top distribution (since $\delta g>0$ with $g_{AB}<0$ relatively much smaller than $\sqrt{g_{AA}g_{BB}}$) around the trap center ($x=0$), which signifies droplet formation~\cite{petrov-ultra-dilute-2016-PhysRevLett.117.100401,bose-gases-low-d-mistakidis}. 
The extent of the flat-top structures is relatively small due to the presence of the trap, a result that has been also showcased before (see, e.g., Refs.~\cite{droplet-harmonic_pot-Mistakidis_2021,ilias-simos-droplet-excitations-PhysRevA.107.023320,ParisiGiorginiMonteCarlo}). 
Moreover, in the absence of impurity--droplet interactions, the impurity exhibits a Gaussian profile localized at the trap center as anticipated for a single particle configuration. 
Note in passing that when comparing these many-body results {with} the eGPE model predictions (not shown), it turns out that the latter does not capture the flat-top region of the majority components but rather yields a Gaussian-like density profile. 
Such deviations become more prominent when impurity--droplet interactions are introduced, as we argue in Section~\ref{sec:Mixed} and present in Figure
~\ref{fig5-assymetric-coupling-plus-eGPE-comp}e--h.

To identify the presence of intra- and intercomponent correlations of our impurity--droplet system, we compute the respective two-body coherence functions~\cite{rdm-Glauber-1999-PhysRevA.59.4595, rdm-sankmann-2008-PhysRevA.78.023615}, which read 
\begin{align}
G^{(2)}_{\sigma \sigma^\prime} (x^{\sigma}_1, x^{\sigma^\prime}_2) = \frac{\rho^{(2)}_{\sigma \sigma^\prime} (x^{\sigma}_1, x^{\sigma^\prime}_2)}{\rho^{(1)}_{\sigma}(x^{\sigma}_1) \rho^{(1)}_{\sigma^\prime}(x^{\sigma^\prime}_2)}.
\end{align}
In this expression, 
 $\rho^{(2)}_{\sigma \sigma^\prime}(x^{\sigma}_1, x^{\sigma^\prime}_2 ) = \bra{\Psi^{\rm MB}}\hat{\rho}^{(2)}_{\sigma\sigma^\prime}\ket{\Psi^{\rm MB}}$ represents  the diagonal of the two-body reduced density matrix involving the two-body reduced density operator   
\begin{equation}
\begin{split}
\hat{\rho}^{(2)}_{\sigma \sigma^\prime} (x^{\sigma}_1, x^{\sigma^\prime}_2 ) = \hat\Psi^{\dagger}_{\sigma} (x^{\sigma}_1) 
\hat\Psi^{\dagger}_{\sigma^\prime} (x^{\sigma^\prime}_2)
\hat\Psi_{\sigma} (x^{\sigma}_1)
\hat\Psi_{\sigma^\prime} (x^{\sigma^\prime}_2).
\end{split}
\end{equation}
 The two-body density matrix describes the probability of simultaneously finding a $\sigma$ and a $\sigma^\prime$ species boson at positions $x_1^{\sigma}$ and  $x_2^{\sigma^\prime}$, respectively. 
Accordingly, a boson of $\sigma$ species and another of $\sigma^\prime$ species feature a bunching (anti-bunching) tendency, if $G^{(2)}_{\sigma \sigma^\prime}(x_1^{\sigma}, x_2^{\sigma^\prime}) > 1$ ($G^{(2)}_{\sigma \sigma^\prime}(x_1^{\sigma}, x_2^{\sigma^\prime}) < 1$), while they are said to be two-body uncorrelated as long as $G^{(2)}_{\sigma \sigma^\prime}(x_1^{\sigma} , x_2^{\sigma^\prime}) = 1$. 
In this sense, for the same (different) species, namely $\sigma=\sigma^{\prime}$ ($\sigma \neq \sigma^{\prime}$), $G^{(2)}_{\sigma \sigma^{\prime}}$ quantifies the spatially resolved intraspecies (interspecies) two-body correlations. 
We finally remark that the two-body coherence function is experimentally accessible via in situ 
 density--density fluctuation measurements \cite{coh2-exp3-Endres2013,coh2-exp1-Tavares_2017,coh2-exp2-Nguyen_2019}.

\begin{figure}
\centering
\includegraphics[width=\linewidth]{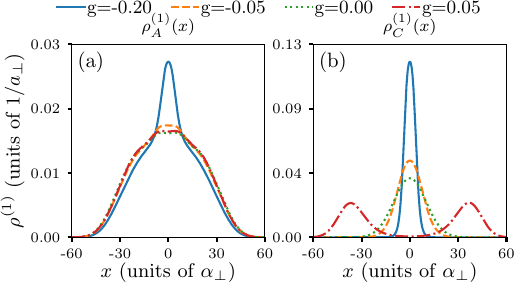}
\caption{Ground$-$state density profiles for (a) the majority species, $\rho^{(1)}_A(x)=\rho^{(1)}_B(x)$, and (b) the impurity, $\rho^{(1)}_C(x)$, for different symmetric impurity{$-$}bath couplings $g_{AC}=g_{BC} \equiv g$ (see legend). Attractive impurity{$-$}bath couplings ($g<0$) lead to impurity localization within the droplet hosts, while repulsive interactions ($g>0$) favor broadening of the impurity distribution and eventual phase separation from the droplet hosts. 
The remaining system parameters are as in Figure~\ref{fig:fig1-decoupled-impurity}.} 
\label{fig:fig2-rho_sym_gAB_-0_02}
\end{figure}

Figure~\ref{fig:fig1-decoupled-impurity}b depicts the intraspecies two-body coherence function, $G^{(2)}_{AA}(x_1^A, x_2^A)$, for the majority component $A$. 
Note that due to the same behavior of $B$ species, the respective $G^{(2)}_{BB}(x_1^B, x_2^B)$ exhibits exactly the same pattern. 
We observe the emergence of anti-bunching between two $A$-species bosons at the same location as can be seen from the diagonal of the coherence function, where $G_{AA}^{(2)}(x_1^A, x_2^A=x_1^A) < 1$.  
Instead, two $A$-species bosons preferentially reside symmetrically on opposite sides with respect to the droplet core being located at the trap center (see $G_{AA}^{(2)}(x_1^A, x_2^A=-x_1^A)>1$).  
This is the characteristic droplet correlation pattern which has been discussed before, see, e.g., Refs.~\cite{ilias-simos-particle-imbalance-PhysRevA.110.023324,ilias-simos-droplet-excitations-PhysRevA.107.023320}. 
Furthermore, it is worth mentioning that the corresponding intercomponent correlation $G^{(2)}_{AB}(x_1^A, x_2^B)$ is somewhat suppressed (see Figure
~\ref{fig:fig3-coherence-symmetric-coupling}b for symmetric finite impurity--droplet couplings) and $G^{(2)}_{AC}(x_1^A, x_2^C) \approx G^{(2)}_{BC}(x_1^B, x_2^C) \approx 1$, since in the current setting $g_{AC}=g_{BC}=0$.

As we shall argue in detail below, introducing finite interactions between the impurity and the droplet hosts has a profound effect on the different components. 
Namely, significant alterations {in the one-body densities of both the droplet bath and the impurity} take place, as the impurity--droplet coupling strength is tuned from the attractive to the repulsive regime.
Interestingly, the intra- and intercomponent correlation character, as captured by the appropriate two-body coherence functions, either maintains the above-described droplet pattern or exhibits deviations from it indicating the development of beyond-LHY phenomenology. 
This analysis will allow us to shed light on the ensuing structural deformations of the above-mentioned two-component droplet behavior caused by the presence of the impurity and its associated dressing by the majority species atoms.

\subsection{Symmetric coupling to the droplet host}\label{sec:Symmetrical}

\begin{figure}
\centering
\includegraphics[width=\linewidth]{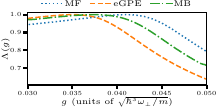}
\caption{Density overlap, $\Lambda(g)$, between the majority components and the impurity (see main text) as a function of the symmetric coupling $g_{AC}=g_{BC} \equiv g\in[0.03, 0.05]$ within the mean{$-$}field (dotted line), the eGPE (dashed line), and the many{$-$}body (dash-dotted line) approaches. 
Correlation effects significantly impact the transition toward impurity{$-$}droplet phase separation, with the LHY correction overestimating the degree of attractive correlations. Other parameters are the same as in Figure~\ref{fig:fig1-decoupled-impurity}.
}
\label{fig:overlap-mf-egpe}
\end{figure}

\begin{figure*}[ht]
\centering
\includegraphics[width=\linewidth]{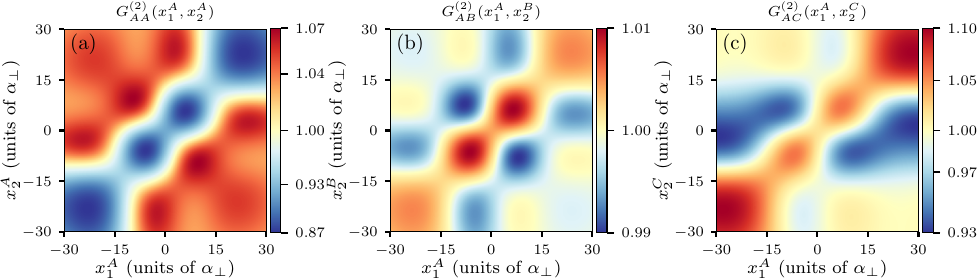}
\caption{
{Two{$-$}body coherence functions of the three{$-$}component system featuring attractive and symmetric impurity{$-$}droplet coupling $g_{AC}=g_{BC} \equiv g=-0.05$. 
(a) Intraspecies two{$-$}body coherence, $G^{(2)}_{AA}(x_1^{A}, x_2^{A})$, of the majority $A$ component. 
Interspecies two{$-$}body coherence, $G^{(2)}_{\sigma \sigma'}(x_1^{\sigma}, x_2^{\sigma'})$, (b)~between the two majority (droplet) components, $\sigma=A$ and $\sigma'=B$, and (c) among the majority $\sigma=A$ and the impurity $\sigma=C$ components. 
The characteristic correlation pattern of the droplet (panel (a)) is maintained in the presence of the impurity, while impurity{$-$}droplet bunching at the same position and anti{$-$}bunching between the impurity and the tails of the droplet occur (panel (c)). 
Other parameters are the same as in Figure~\ref{fig:fig1-decoupled-impurity}.}
}
\label{fig:fig3-coherence-symmetric-coupling}
\end{figure*}

To depart from the non-interacting impurity scenario discussed above, we first consider a system in which the impurity species $C$ is symmetrically coupled to the $A$, $B$ majority components, such that $g_{AC}=g_{BC} \equiv g$. 
The latter condition {combined} with  
$g_{AA}=g_{BB}$, $m_A=m_B$, and $N_A=N_B$ (as in Section~\ref{sec:Decoupled}) ensure that the two majority components exhibit identical droplet density distributions, i.e., $\rho_A^{(1)}(x) = \rho_B^{(1)}(x)$. 
Hence, in what follows, we discuss only one of them and focus on the impact of the impurity--droplet coupling on both the droplet and the impurity configurations.

Figure~\ref{fig:fig2-rho_sym_gAB_-0_02} shows the emergent one-body densities of the majority component $A$ and the impurity $C$ for different impurity--droplet interactions ($g$) ranging from attractive to non-interacting and eventually to repulsive values. 
It becomes apparent that at stronger attractive couplings, e.g., $g=-0.2$, 
the droplet hosts (Figure~\ref{fig:fig2-rho_sym_gAB_-0_02}a) develop a pronounced hump at the overlap region with the impurity, while the impurity resides in the vicinity of the trap center, featuring a highly localized density distribution (Figure~\ref{fig:fig2-rho_sym_gAB_-0_02}b).
This is attributed to the attractive impurity--droplet coupling, $g<0$, which enforces a certain interaction dependent amount of the droplet atoms {accumulating} close to the impurity. 
This backaction of the impurity to the droplet host designates the dressing of the former by the latter, and it is reminiscent of the behavior observed {for an} attractive well {acting} on the droplet~\cite{Bristy_defect_drops}, {as well as that of an} attractively interacting impurity with a repulsive Bose gas, {which gives} rise to an attractive Bose polaron~\cite{Grusdt_2025,mistakidis2020many} (see also Appendix~\ref{app:dressing}).

The interaction-dependent nature of the above-described process at $g<0$ is evident upon reducing the attractive impurity--droplet interaction, e.g., to  $g=-0.05$. 
As a consequence, the density hump building upon the droplet components (around $x=0$) becomes less pronounced, gradually transitioning to a flat-top structure. Simultaneously, the impurity exhibits spatial delocalization deforming into a broader Gaussian profile {while} remaining within the droplet extent (see also the discussion below for the spatial overlap among the impurity and the droplet). 
This indicates the tendency of the impurity to undress and decouple from the droplet.  Eventually, at $g=0$, we retrieve the distributions discussed in Section~\ref{sec:Decoupled}  corresponding to a flat-top density for the droplet and a broader Gaussian for the impurity (see also Figure~\ref{fig:fig1-decoupled-impurity}a). 
It is worth mentioning here that the interaction range, within which the shape of the majority components is almost unperturbed, depends on the presence of the external trap. 
The latter enforces smaller length scale structures which have simultaneously higher amplitude. 
Hence, relatively larger interactions are required to deform the majority components compared to the free space scenario where the droplet hosts are more elongated and have a  smaller amplitude (see also Appendix~\ref{app:impurity:freespace}). 
Turning to weak repulsive interactions, e.g., $g=0.05<g_{AA}$, we observe that the impurity is expelled from the droplet bath, splitting into two symmetric density distributions.
These are primarily located outside the droplet of the majority components and have a relatively small overlap with the droplet tails due to the weak repulsion and the presence of the harmonic trap. 
A similar early phase-separation process between a third component and a two-component droplet has been recently discussed within the eGPE framework in Ref.~\cite{ilias-simos-SciPostPhys.19.5.133}.
On the other hand, the droplet density is almost unperturbed, exhibiting a similar flat-top structure as in the non-interacting ($g=0$) case (see Figure~\ref{fig:fig2-rho_sym_gAB_-0_02}a). 

To gain further insights into the emerging phase-separation process between the impurity and the majority (droplet) host, we next  evaluate their density overlap, $\Lambda(g)=\int dx \sqrt{\rho_A^{(1)}(x) \rho_C^{(1)}(x)}$ for different impurity--droplet repulsive couplings. 
The lower bound of this quantity is $\Lambda(g)=0$, indicating that $\rho_A^{(1)}(x)$ and $\rho_C^{(1)}(x)$ are non-overlapping {and therefore fully phase-separated (immiscible)}. 
{On the other hand,} the upper bound {of this measure} is $\Lambda(g)=1$, which is achieved as long as $\rho_A^{(1)}(x)=\rho_C^{(1)}(x)$ holds, {implying complete overlap (miscibility) between the impurity and the droplet hosts. For $0<\Lambda(g)<1$, partial phase-separation takes place between the involved components, which is the case for the considered interaction strengths presented in Figure~\ref{fig:overlap-mf-egpe}.} 
This measure is demonstrated in Figure~\ref{fig:overlap-mf-egpe} within three different methods, namely the mean-field approximation (Equation~(\ref{3comp_eGPEs}) with $\mathcal{E}_{LHY}=0$), where all correlations are ignored, the eGPEs (Equation~(\ref{3comp_eGPEs})) accounting for droplet correlations to first-order via the LHY term, and the full many-body ML-MCTDHX method (Equation~(\ref{eq:psi_mlx})) incorporating all correlations. 
We focus on the parametric range of weakly repulsive couplings, $g\in[0.03, 0.05]$, where the transition towards the {partially} phase-separated impurity--droplet configurations is found to occur within all three different approaches. 
As can be seen, in all three methods, the overlap $\Lambda(g) \approx 1$ for $g\in[0.03, 0.035]$, implying that the impurity is almost  fully immersed in the droplet bath. 
In particular, within this interaction range, the impurity is localized within the droplet bath and as $g \to 0.035$ the impurity distribution broadens and its peak amplitude decreases, gradually approaching the shape of the bath density profile. 
For this reason, $\Lambda(g)$ shows a slightly increasing tendency approaching unity. 
However, for further increasing interactions, $\Lambda(g)$ decays significantly from unity, signaling that the impurity starts {being} expelled from its droplet host.
Interestingly, this gradual phase-separation behavior depends on the level of correlations. 
Namely, in the mean-field approximation where all correlations are absent, the impurity begins to phase-separate at  $g\approx0.043$.
Taking into account quantum fluctuations of the droplet to first-order in perturbation theory, modeled by the LHY term in the eGPEs, expedites the splitting mechanism, which is found to begin at $g\approx0.037$. 
This difference appears to be primarily due to the attractive nature of the LHY term, which results in bath configurations featuring somewhat smaller (larger) spatial extents (density amplitudes), as compared to the mean-field predictions. 
This, in turn, results in the impurity experiencing a stronger repulsion from the bath $\propto 2 g |\Psi_A|^2$, within the eGPE approach, facilitating its expulsion.

Finally, in the full many-body treatment, the impurity remains trapped within the droplet components up to $g\approx0.04$. 
This indicates that the eGPEs slightly overestimate the attractive impact of the correlations, owing to higher order effects, not accounted for in the model of Equation~\eqref{3comp_eGPEs}. 
Such deviations manifesting in droplet density profiles  have been previously reported by comparing eGPE and ab initio simulations~\cite{ilias-simos-droplet-excitations-PhysRevA.107.023320,ParisiGiorginiMonteCarlo}. 
The aforementioned corrections may include beyond LHY correlations in the bath, the impact of the harmonic trap to the correlations, and the intercomponent anti-correlations between the bath and the impurity (c.f. Figure~\ref{fig:fig3-coherence-symmetric-coupling}c). 
Still, it appears that in the presence of all correlations, the expulsion of the impurity from the bath is favored as compared to the mean-field  approximation. 
{This result may be important for subsequent studies} focusing on the impurity's polaronic properties and it is in line with previous results demonstrating that the phase-separation threshold occurs at relatively smaller intercomponent repulsions in the presence of correlations~\cite{correl-effects-2species-Mistakidis_2018,Pyzh}.

Next, we explore the underlying correlation patterns of the impurity--droplet symmetrically coupled setting. 
As an example, we focus on the case of a weakly attractive impurity, e.g., for $g=-0.05$, presented in Figure~\ref{fig:fig3-coherence-symmetric-coupling}. 
Specifically, the majority (droplet) component retains the above-described (see also Figure~\ref{fig:fig1-decoupled-impurity}b and Section~\ref{sec:Decoupled}) intraspecies correlation pattern. 
The latter is characterized by anti-bunching and bunching across the droplet core (diagonal of $G^{(2)}_{AA}(x_1^A, x_2^A)<1$ in Figure~\ref{fig:fig3-coherence-symmetric-coupling}a) and between symmetric positions with respect to the trap center (see the anti-diagonal of $G^{(2)}_{AA}(x_1^A, x_2^A)>1$ in Figure~\ref{fig:fig3-coherence-symmetric-coupling}a), respectively.  
A pattern similar to the above-described pattern takes place in the interspecies coherence between the majority components, i.e., $G^{(2)}_{AB}(x_1^A, x_2^B)$ (see Figure~\ref{fig:fig3-coherence-symmetric-coupling}b). 
Indeed, a boson of species $A$ and another of species $B$ feature a weak bunching (anti-bunching) tendency at the same (different) locations as shown by the diagonal (anti-diagonal) of the interspecies coherence, where $G^{(2)}_{AB}(x_1^A, x_2^B=x_1^A)>1$ ($G^{(2)}_{AB}(x_1^A, x_2^B=-x_1^A)<1$) as depicted in Figure \ref{fig:fig3-coherence-symmetric-coupling}b.  
Noticeably, the small deviation from unity of $G^{(2)}_{AB}(x_1^A, x_2^B)$ suggests a nearly product state among the majority components. 
Hence, entanglement between the droplet components is found to be suppressed~\cite{dmrg-QD-dimerized-QD-Astrakharchik-PhysRevLett.126.023001, dmrg-QD-particle-imbalance-Astrakharchik-10.21468/SciPostPhys.16.3.074, ParisiMonteCarlo2019, ParisiGiorginiMonteCarlo}.  
It is worth mentioning that the intraspecies correlation structure of each of the majority components for stronger attractive $g$ values is significantly altered, as we discuss later on in Section~\ref{sec:Mixed} (see also Figure
~\ref{fig:fig5-2b-coh-bath-mixed}a). 
This means that the impurity may structurally deform the correlation pattern of its host in a non-trivial manner.  

Meanwhile, the interspecies impurity--droplet coherence signifies the presence of finite impurity--droplet entanglement. 
Indeed, it exhibits a bunching trend across its diagonal (see $G^{(2)}_{AC}(x_1^A, x_2^C=x_1^A)>1$ in Figure~\ref{fig:fig3-coherence-symmetric-coupling}c). 
This is attributed to the attractive impurity--droplet coupling ($g<0$), promoting the binding of a certain amount of majority atoms to the impurity reflected on the droplet distribution by the manifestation of a density hump (see also Figure~\ref{fig:fig2-rho_sym_gAB_-0_02}a).  
In addition, an anti-bunching behavior occurs between the impurity residing near the trap center and a boson of the majority component that lies outside the droplet's flat-top region (see, e.g., $G^{(2)}_{AC}(x_1^A \approx 20, x_2^C\approx 0 )<1$ in Figure \ref{fig:fig3-coherence-symmetric-coupling}c).

Overall, we can conclude that the majority components tend to retain their droplet character, in spite of the ensuing structural deformations caused by the presence of the impurity. 
This is indicative of the robustness of the droplet configurations against the perturbing impurity. 
The latter acts as a knob not only for altering the droplet distribution, but also for changing the intercomponent correlation patterns of the impurity and the bath components, hinting to modifications of the impurity--droplet entanglement. 

\begin{figure*}
\centering
\includegraphics[width=\linewidth]{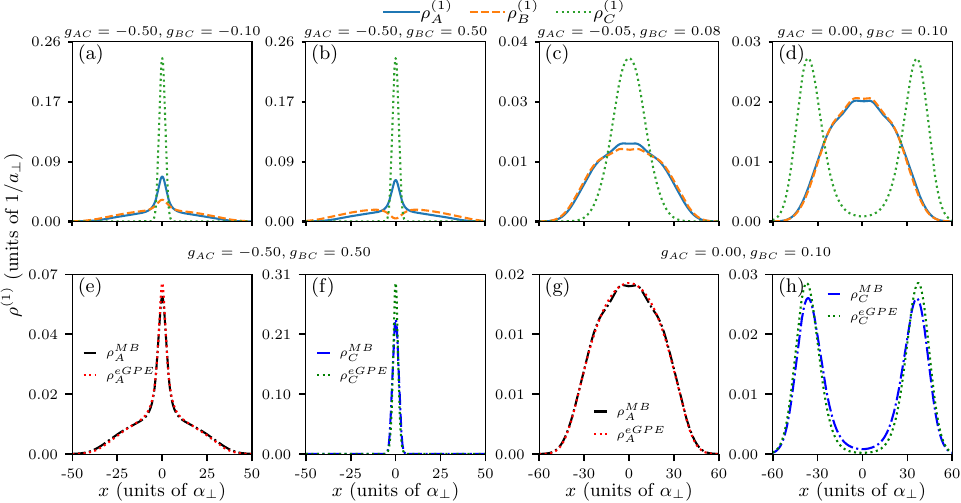}
\caption{{(a{$-$}d) Density distributions of the three{$-$}component mixture for representative mixed impurity{$-$}bath couplings $g_{AC} \neq g_{BC}$ (see legend), obtained within the many{$-$}body method. 
Strong attractive{$-$}repulsive mixed couplings lead to a pronounced hump{$-$}dip alternating structure in the majority components (panel (b)), while weaker interactions result in nearly flat{$-$}top configurations (panel (c)). Eventually,  phase{$-$}separation occurs when the impurity is decoupled with one majority component and experiences repulsion from the other (panel (d)). 
(e{$-$}h) Comparison of the different component density profiles as predicted by the many{$-$}body and the eGPE approaches (see legends) for (e,f) $g_{AC}=-g_{BC}=-0.5$ and (g,h) $g_{AC}=0$, $g_{BC}=0.1$. 
The eGPE captures the overall structure but overestimates the density peaks and impurity localization. 
For simplicity, the majority component $B$ is omitted since it shows a similar deviation between the two methods as component $A$.  
Other system parameters are as in Figure~\ref{fig:fig1-decoupled-impurity}. }
}
\label{fig5-assymetric-coupling-plus-eGPE-comp}
\end{figure*}

\subsection{Mixed impurity-droplet couplings}\label{sec:Mixed}

In the following, we extend our considerations to the arguably more general cases of mixed couplings between the majority components and the single impurity.
Namely, $g_{AC} \neq g_{BC}$, referring to either opposite or same interaction signs of distinct magnitude. 
This interplay of interactions breaks the previously discussed symmetry between the majority components, which are thus now well distinguishable, yielding modified two-component droplet structures. 
Below, we focus on characteristic impurity--droplet interaction combinations and explicate the emergent features of the three-component system. 
We first address the impact of relatively strong coupling strengths, $|g_{\sigma C}| \geq g_{\sigma\sigma}$.  
Thereafter, we discuss the relevant phenomenology at weaker interactions closer to the decoupled limit, i.e., $|g_{\sigma C}| \leq g_{\sigma\sigma}$, with $\sigma = A,B$.

Our starting point is the system where both impurity--bath couplings are somewhat strongly attractive but of different strength, namely $g_{AC}=-0.5$ and $g_{BC}=-0.1$. 
Here, the total energy of the system can be more flexibly tuned by adjusting the impurity--bath couplings as compared to the corresponding case of symmetric couplings. 
The resultant ground-state density profiles of all components are presented in Figure~\ref{fig5-assymetric-coupling-plus-eGPE-comp}a. 
As expected, the impurity--bath attraction imprints characteristic density humps on top of both majority component densities near the location of the impurity. 
This is traced back to the emergent binding between the impurity and the atoms of the majority components which is naturally more prominent for stronger interactions. 
This feature can be directly seen by inspecting the underlying interaction energies (not shown). 
The aforementioned interaction driven binding strength is reflected by the more pronounced density peak in component $A$ as compared to component $B$, implying that more atoms of component $A$ are bound to the impurity. 
The remaining atoms of the majority components $A$, $B$ reside in the elongated density tails of the relevant distributions, mainly featuring binding between each other due to $g_{AB}<0$. 
As expected, the impurity is highly localized near the trap center ($x=0$), similarly to the earlier studied symmetrically coupled cases (see Section~\ref{sec:Symmetrical} and Figure~\ref{fig:fig2-rho_sym_gAB_-0_02}). 
We also remark that the degree of spatial localization of the impurity is dictated by the strength of the attractive impurity--bath couplings, as will be discussed below.

Next, we turn our attention to different sign impurity--bath couplings.
Namely, the case where the impurity of species $C$ still interacts attractively with the majority component $A$ with strength $g_{AC}=-0.5$, while it couples repulsively to the majority component $B$ with strength $g_{BC}=0.5$ (see Figure~\ref{fig5-assymetric-coupling-plus-eGPE-comp}b).  
The impurity is localized in the vicinity of the trap center with a peak amplitude slightly smaller compared to the previous scenario with $g_{BC}=-0.1$ (Figure~\ref{fig5-assymetric-coupling-plus-eGPE-comp}a) due to the repulsive $g_{BC}=0.5$. 
Accordingly, the attractive $g_{AC}$ value is responsible for the density hump appearing in $\rho^{(1)}_A(x)$, facilitating the accumulation of type $A$ atoms near the impurity, which remains trapped by component $A$.   
In contrast, the repulsive $g_{BC}$ strength imprints a prominent density dip to $\rho^{(1)}_B(x)$, since $B$ species atoms are pushed away from the impurity and tend to phase-separate from the latter. 
Here, the impurity lies around the trap center due to its strong binding with component $A$ and thus component $B$ splits into two symmetric humps. {It is thus the dominance of the attractive interaction between the impurity and component $A$ that preserves the localization of the impurity.}
Hence, it is possible to configure alternating miscible and immiscible configurations of the impurity with the distinct droplet components. 
This result is expected to be of interest for forthcoming investigations of dressed quasi-particles, since it can affect the residue and lifetime of such states~\cite{Grusdt_2025}. 
Besides impurity physics, we anticipate that this phenomenology persists for larger atom numbers in the third component and can trigger studies on dynamical miscibility-immiscibility phase transitions in a similar vein to previous investigations in binary systems~\cite{Tojo,correl-effects-2species-Mistakidis_2018}. 

Tuning the impurity--bath couplings toward the weak interaction regime, we observe drastically altered majority component distributions, reminiscent of the corresponding scenario of symmetric couplings outlined in Section~\ref{sec:Symmetrical} and illustrated in Figure~\ref{fig:fig2-rho_sym_gAB_-0_02}a. 
A prototypical example of weak mixed couplings is depicted in Figure~\ref{fig5-assymetric-coupling-plus-eGPE-comp}c, where the impurity experiences attraction (repulsion) with component $A$ ($B$) of strength $g_{AC}=-0.05$ \mbox{($g_{BC}=0.08$)}. 
Evidently, due to the weak impurity--droplet interactions, both majority components exhibit a nearly flat-top density profile centered around $x=0$. 
Notice here that the density amplitude at the flat-top region appears to be slightly smaller for the repulsively coupled component $B$ as compared to the attractively interacting component $A$. 
Accordingly, the distribution of component $B$ is slightly broader with respect to the one of $A$. 
Both of these aspects are a direct consequence of the different impurity--bath interaction signs and the fact that $|g_{AC}|$ is slightly smaller than $|g_{BC}|$. 
Furthermore, as a result of the weak interactions, the impurity distribution is again of Gaussian type but it is relatively broader than the stronger (in magnitude) interacting cases (see Figure~\ref{fig5-assymetric-coupling-plus-eGPE-comp}b,c). 
Here, it is again the attractive $g_{AC}$ together with the weakly repulsive $g_{BC}$ that facilitate the impurity localization at the trap center.

For completeness, we also explore the situation where the impurity is decoupled  {from} one of the majority components ($g_{AC}=0$) and features repulsion by the other one ($g_{BC}=0.1$) (see Figure~\ref{fig5-assymetric-coupling-plus-eGPE-comp}d).  
Here, both majority components maintain their flat-top profile around the trap center, with the impurity becoming phase-separated from them by splitting into two symmetrically placed configurations, due to the repulsive $g_{BC} \geq g_{AA}$, surrounding the majority components. 
It is indeed energetically preferential for the minority component to split into two parts, which lie outside the majority components, due to the larger atom number accommodated in the latter.
This phase-separation mechanism is similar to the one which was previously discussed for symmetric weak impurity--droplet repulsions (see also Section~\ref{sec:Symmetrical}).     

Having discussed characteristic configurations of the droplet hosts and the impurity at different mixed coupling strengths within the many-body approach, we subsequently provide comparisons with the predictions of the eGPE model of Equation~(\ref{3comp_eGPEs}). 
This will allow us to infer the role of beyond-LHY correlations in the ensuing three-component phases at the considered mesoscopic atom number regime. 
As a representative mixed-interaction setting, we invoke the one with strong attractive and repulsive interactions between the impurity and the different bath components, namely $g_{AC}=-0.5$ and $g_{BC}=0.5$. 
The underlying $A$-species majority and $C$-species impurity density distributions are shown in Figure~\ref{fig5-assymetric-coupling-plus-eGPE-comp}e and d, respectively, within the many-body and the eGPE methods. 
The agreement between the predictions of the two methods is adequate, although the eGPE approach tends to overestimate the maximum density of both components and especially of the impurity. 
{In addition, the spatial localization of the impurity is slightly reduced in the many-body case as compared to the eGPE prediction. This can be understood as follows. 
The droplet host has a slightly broader density profile in the presence of beyond-LHY correlations as it has been also shown in Refs.~\cite{ParisiGiorginiMonteCarlo,droplet-harmonic_pot-Mistakidis_2021}. 
In turn, the droplet density acts as an effective potential to the impurity, which as a consequence experiences a somewhat less localization tendency in the presence of beyond-LHY correlations.}
Note also that the eGPE approach captures only qualitatively the density dip arising in the $B$-species majority component (not shown for brevity). 

Subsequently, we compare the two methods at the phase-separation regime (see also Figure~\ref{fig5-assymetric-coupling-plus-eGPE-comp}d) realized here by the repulsively interacting (non-interacting) minority with majority component $B$ ($A$). 
The comparison of the resulting density distribution of the majority $A$ and minority $C$ components is presented in Figure \ref{fig5-assymetric-coupling-plus-eGPE-comp}g,h. 
As can be seen, the eGPE approach cannot  
accurately capture the flat-top profile of the majority components (Figure~\ref{fig5-assymetric-coupling-plus-eGPE-comp}g) but rather leads to a Gaussian-like distribution. This is attributed to the presence of the harmonic trap. 
Such an effect, where signatures of a flat-top density profile are predicted in the many-body approach,  but not within the eGPE method in the presence of a weak harmonic trap, has been previously reported for symmetric droplet structures (see e.g., Ref.~\cite{ilias-simos-droplet-excitations-PhysRevA.107.023320}). 
Moreover, the impurity's distribution exhibits spatial separation with the majority components in both methods (see Figure~\ref{fig5-assymetric-coupling-plus-eGPE-comp}h). 
Here, again the eGPE prediction is qualitatively accurate but the impurity's density peak and width are somewhat overestimated when compared to the ones obtained within the many-body method.   

\begin{figure}
\centering
\includegraphics[width=\linewidth]{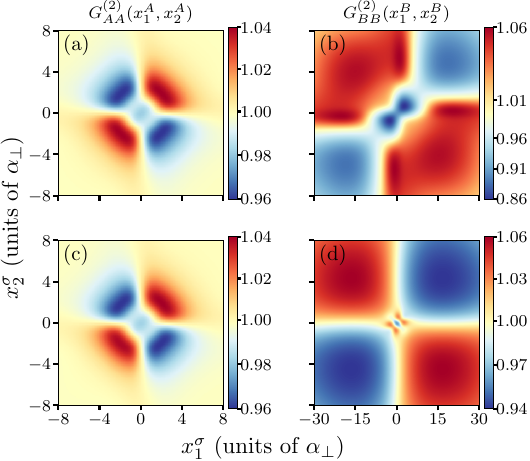}
\caption{{Two{$-$}body intracomponent coherence functions of the majority (a,c) $A$ and (b,d) $B$ components for mixed impurity{$-$}bath couplings (a,b) $g_{AC}=-0.5$, $g_{BC}=-0.1$ and (c,d) $g_{AC}=-0.5$, $g_{BC}=0.5$. 
A transition from bunching{$-$}to{$-$}anti{$-$}bunching (across the diagonal) for component $A$ occurs, distracting the characteristic droplet correlation pattern. 
Similarly, a strong $g_{BC}>0$ results in the modification of $G^{(2)}_{BB}(x_1^B,x_2^B)$, as can be seen by inspecting its diagonal.  
The remaining parameters are as in Figure~\ref{fig:fig1-decoupled-impurity}.}}
\label{fig:fig5-2b-coh-bath-mixed}
\end{figure}

Our next task is to analyze the associated correlation patterns for characteristic three-component configurations appearing at mixed impurity--droplet couplings.  
Below, our focus is put on the intraspecies correlations of the bath components in order to argue whether the impurity can be utilized as a knob to manipulate the correlation properties of the droplet host. 

For this purpose, we employ two different settings. 
In the first system, both majority components are attractively coupled to the impurity via $g_{AC}=-0.5$, $ g_{BC}=-0.1$. 
Here, the majority components exhibit a density hump at the vicinity of the impurity (see also Figure~\ref{fig5-assymetric-coupling-plus-eGPE-comp}a for the respective one-body density distributions). 
The two-body coherence functions of the majority components $A$ and $B$ are illustrated in Figure~\ref{fig:fig5-2b-coh-bath-mixed}a,b. 
As can be seen, the correlation pattern of the more strongly coupled component, $A$, is significantly altered from the characteristic droplet pattern. 
Namely, along the main diagonal, there is a transition from bunching behavior, for spatial regions outside the density hump, towards a weakly anti-bunching tendency within the region of the density hump located near the trap center. 
Note in passing that this correlation behavior of the bath can also occur in the case of symmetric and sufficiently strong attractive impurity--droplet coupling (not shown). 
On the other hand, the correlation pattern of the majority component $B$ 
(Figure~\ref{fig:fig5-2b-coh-bath-mixed}b) retains its droplet-like character. 
It exhibits anti-bunching along the diagonal, implying that two bosons do not prefer to lie at the same position in the droplet, and a bunching tendency across the off-diagonal, which means that two bosons are likely to be symmetrically placed with respect to the droplet core.

\begin{figure*}
\centering
\includegraphics[width=\linewidth]{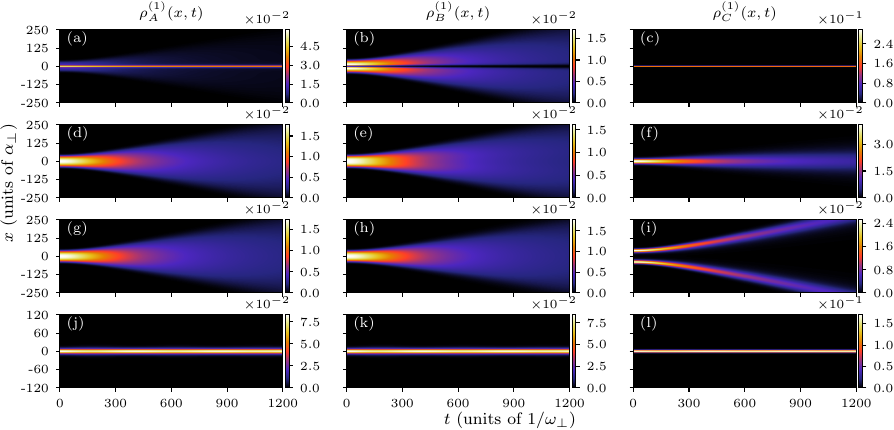}
\caption{
Time{$-$}evolution of the $\sigma=A,B,C$ species one{$-$}body density, $\rho^{(1)}_{\sigma}(x,t)$, within the eGPE approach, after trap release at $t=0$ from $\omega=0.005$ to $\omega=0$. 
Panels correspond to (a{$-$}c) $(g_{AB},g_{AC},g_{BC}) = (-0.02,-0.5, 0.5)$, 
(d{$-$}f) $(g_{AB},g_{AC},g_{BC}) = (-0.02,-0.05, 0.08)$,   
(g{$-$}i) $(g_{AB},g_{AC},g_{BC}) = (-0.02,0.0,0.1)$, and (j{$-$}l)  $(g_{AB},g_{AC},g_{BC}) = (-0.1,-0.05, -0.08)$. 
In all cases, the first (second) column illustrates the dynamics of the majority component $A$ ($B$), and the third visualizes the impurity species $C$. 
For mixed strong attractive{$-$}repulsive impurity{$-$}bath interactions (panels (a{$-$}c)), the majority component  shows expansion, while effectively trapping the other components sustaining their shape during the evolution. 
Expansion occurs in the cases featuring weak intercomponent  attraction. 
Other system parameters are as in Figure~\ref{fig:fig1-decoupled-impurity}. }
\label{fig:eGPE-dyn}
\end{figure*}

The second system that we analyze refers to opposite sign couplings between the bath components and the impurity, namely $g_{AC}=-0.5$, and $ g_{BC}=0.5$. 
Their density distributions are shown in Figure~\ref{fig5-assymetric-coupling-plus-eGPE-comp}b and the intraspecies coherence functions of the bath are depicted in Figure~\ref{fig:fig5-2b-coh-bath-mixed}c,d. 
The correlation pattern of the attractively coupled component, $A$, to the impurity maintains the same pattern as in the previously analyzed case presented in Figure~\ref{fig:fig5-2b-coh-bath-mixed}a.  
In contrast, the repulsively coupled component $B$ preserves two-body coherence reminiscent of the droplet, except within the spatial region of the density dip, where we observe an intriguing bunching-to-anti-bunching-to-bunching transition (see the region $x' \in [-5,5]$ and $x \in [-5, 5]$ in Figure~\ref{fig:fig5-2b-coh-bath-mixed}d).
We note, finally, that the two-body interspecies coherence between either of the bath components and the impurity is qualitatively similar to the one we presented in Figure~\ref{fig:fig3-coherence-symmetric-coupling}c for the symmetric coupling case. 
Namely, the diagonal shows a bunching (anti-bunching) behavior for attractive (repulsive) couplings (not shown). 
This supports the existence of finite impurity--bath entanglement, which is crucial for quasi-particle formation~\cite{Grusdt_2025,bose-gases-low-d-mistakidis}, and also hints toward beyond-LHY contributions since the LHY term does not account for intercomponent correlation processes~\cite{petrov-ultra-dilute-2016-PhysRevLett.117.100401,ilias-simos-SciPostPhys.19.5.133}.

\section{Expansion Dynamics}\label{sec:dynamics} 

To shed light on the dynamical response of the above-discussed three-component configurations, we finally monitor their time-evolution following a sudden removal of the external harmonic trap at $t=0$. 
Such a protocol is routinely implemented in corresponding experiments via time-of-flight imaging~\cite{Cavicchioli,CabreraTarruellDropExp}.  
Since the predictions of the LHY approximation showcase fairly good agreement with the results obtained within the full many-body approach on the ground-state level, for simplicity, we use this framework to study the resulting dynamics. 
Our discussion, below, pertains to selected configurations whose ground-state properties have been analyzed above. 
This investigation is expected to motivate future studies on the full quantum dynamics for exploring the interplay of correlations but also the rise of beyond-LHY contributions.

We first focus on strongly mixed interactions, e.g., $(g_{AC},g_{BC})=(-0.5,0.5)$. 
Here, the impurity is tightly bound to the majority component $A$, while simultaneously repelling atoms of component $B$(see also Figure~\ref{fig5-assymetric-coupling-plus-eGPE-comp}b for the ground-state density profiles). 
After the trap removal, component $A$ retains its central core (Figure~\ref{fig:eGPE-dyn}a), which is strongly bound with the impurity, while the remaining unbound atoms residing at the tails of $\rho_A^{(1)}(x,t)$ feature expansion. 
Instead, the two density fragments of the repulsively coupled component $B$ undergo a noticeable expansion while maintaining the central dip, which broadens {during} the evolution (Figure~\ref{fig:eGPE-dyn}b), partly due to the aforementioned expansion of the tails of component $A$. 
This structure of component $B$ essentially provides an effective trapping potential at the center for both the impurity and the bound fraction of component $A$. 
As such, the impurity remains almost intact during the dynamics due to this effective potential and its binding with component $A$ (see Figure~\ref{fig:eGPE-dyn}c).

A somewhat altered dynamical response is observed for weak mixed interactions, such as $(g_{AC},g_{BC})=(-0.05,0.08)$ (see Figure~\ref{fig:eGPE-dyn}d--f). 
Due to the weak impurity--bath couplings, the impurity is unable to significantly bind atoms from the majority components or deform their ground-state configuration (see also Figure~\ref{fig5-assymetric-coupling-plus-eGPE-comp}c). 
Removing the trap entails an overall expansion of all three species. 
The majority components $A$ and $B$ spread symmetrically and at almost the same rate (Figure~\ref{3comp_eGPEs}d,e). 
The impurity experiences a similar expansion but at a substantially smaller rate (Figure~\ref{3comp_eGPEs}f), remaining comparatively more localized than the majority components.
This behavior demonstrates that, in the weak-coupling regime, the initially trapped configurations are not sustained as bound structures, and the expansion of the different species is governed by the released kinetic pressure. 
This is in part due to the fact that self-bound configurations in free space exhibit significantly larger length scales for weak interactions (see also Appendix~\ref{app:impurity:freespace}).

Next, we explore the dynamics of partially decoupled configurations, where for instance $(g_{AC},g_{BC})=(0,0.1)$. 
Here, the impurity does not interact with component $A$ and is expelled by component $B$, facilitating their phase-separation (see Figure~\ref{fig5-assymetric-coupling-plus-eGPE-comp}d). 
The emergent dynamics after a quench to $\omega=0$ is presented in Figure~\ref{fig:eGPE-dyn}g--i.  
As can be seen, both of the majority components feature a clear expansion tendency (Figure~\ref{fig:eGPE-dyn}g,h) with the spreading rate of component $A$ being slightly faster than the one of component $B$. 
This is attributed to the repulsive interaction of component $B$ with the impurity, which effectively acts as a barrier slowing down the initial expansion of component $B$. 
Accordingly, the two fragments of the impurity (in the absence of binding with the majority components) expand  outward.

Concluding, the above results reveal that the dynamical response of the considered three-component system is sensitive to both the sign and the magnitude of the impurity--bath interactions. 
Strong attractive interactions between the impurity and one of the bath components can partially retain the stability of the mixture against expansion. 
In contrast, weak attractive or repulsive impurity--bath couplings favor spreading of the involved components. 
The role of attractive intercomponent interactions acting  against expansion of the involved components is further corroborated by the response of the setup shown in Figure~\ref{fig:eGPE-dyn}j--l.  
Here, the two-component majority subsystem lies in the so-called LHY fluid regime~\cite{Skov_fluid}, where $g_{AA}g_{BB}-g_{AB}^2=0$. 
Hence, in the absence of impurity--bath coupling (i.e., $g_{\sigma C}=0$ in Equation~(\ref{eGPE_major})), mean-field interactions cancel out leaving the majority species subsystem to solely experience quantum fluctuations.  
Additionally, we assume attractive impurity--bath interactions $(g_{AC},g_{BC}) = (-0.05, -0.08)$. 
The dynamics of this system following a trap release unveils that all components remain intact as time evolves, thereby preserving their original shape. 
This behavior is inherently related to the attractive intercomponent interactions, since setting one of these couplings either to zero or to the repulsive regime yields expansion of the underlying components (not shown for brevity). 

\section{Summary and Outlook}\label{sec:conclusions}

We have investigated the ground-state properties and dynamical response of a single impurity embedded into a one-dimensional two-component quantum droplet. 
Our analysis is based on ab initio many-body simulations complemented by the predictions of the suitable coupled  extended Gross--Pitaevskii model. 
It is demonstrated that when tuning the impurity--droplet interactions, the presence of the impurity allows the control of the distribution of the droplet hosts and in particular selectively deforms their density distributions and accompanying correlation structures. 
This phenomenology is uncovered by considering different impurity--droplet couplings ranging from symmetric (i.e., the same) to mixed (different magnitude and/or sign) ones.

Specifically, for attractive impurity--droplet couplings, it is shown that the impurity distribution is spatially localized and remains within the droplet core. 
Importantly, it induces a pronounced density hump in the droplet components due to the induced binding of majority species atoms around it, thus reflecting the formation of a dressed impurity state. 
In contrast, repulsive impurity--droplet interactions result in impurity expulsion and the gradual emergence of phase separation between the impurity and the droplet. 
Accordingly, mixed impurity--bath couplings lead to selective modifications of the droplet distributions, giving rise to more complex three-component configurations featuring localized density humps and dips in the individual majority components.

Inspecting the intracomponent two-body coherence functions of the droplet majority components, it is found that in the weak coupling regime the droplet retains its characteristic correlation pattern. 
The latter consists of an anti-bunching tendency of two bosons located at the same position across the droplet and a bunching behavior for two bosons symmetrically placed with respect to the droplet core. 
Instead, stronger impurity--droplet couplings modify the above-mentioned correlation structure leading to intriguing characteristics such as a transition from bunching-to-anti-bunching at the droplet peak. 
This entails the ability of the impurity to affect the correlation patterns of droplets. 
Turning to impurity--droplet intercomponent two-body coherence functions, we unveil a general bunching (anti-bunching) trend at the same (different) positions confirming the presence of impurity--droplet entanglement and supporting the impurity's dressing by the majority components.

A comparison between our ab initio simulations and the predictions of the eGPE approach reveals that the latter is capable to overall capture the aforementioned impurity induced droplet deformations and the phase-separation processes. 
However, the eGPE model systematically overestimates the spatial localization of the impurity and the peak density of the majority components, while it predicts an earlier onset of phase-separation. 
These discrepancies manifest the involvement of beyond-LHY correlations, with the latter supporting the ability of the droplet hosts to trap the impurity. 

Finally, we examined the emergent dynamical response of the three-component system after a trap release. 
We showcase that the sign and the magnitude of the intercomponent interactions between the impurity and the hosts regulates the ensuing dynamics of the individual components. 
Indeed, strong attractive interactions are able to sustain the bound atom fraction of the respective  majority component. 
On the other hand, for weak attractive or repulsive couplings all components feature a prominent expansion whose rate is enhanced for stronger repulsive interactions.  

There are several interesting extensions based on our results which open the door for exploring impurity physics and the presence of quantum fluctuations in droplet states of matter. 
As a first step, it is worthwhile to perform a systematic investigation of the underlying phase diagram with respect to the different intra- and inter-species interactions elucidating in particular the role of impurity--droplet entanglement.  
In this direction, it is intriguing to explore the non-equilibrium quantum dynamics of our impurity setting following interaction quenches across different phases.
In particular, it would be compelling to unfold pattern formation that competes with impurity-regulated modulational instability~\cite{mithun2020modulational} and to assess the possible emergence of rogue wave structures recently identified in droplet environments~\cite{Chandramouli_RWs}.  
The generalization of our results to higher dimensions is certainly desirable. 
Another possibility is to emulate a radiofrequency spectroscopy scheme by utilizing a spinor impurity in order to realize dressed polaron states in droplet hosts, thereby enabling the characterization of quasiparticle related properties such as their residue and effective mass. 
Finally, the study of two-impurities either of bosonic or fermionic nature will allow the investigation of the effect of induced interactions and long-range correlations mediated by the droplet, which is a topic of growing interest.  
\section*{Acknowledgements}

S.I.M acknowledges support by the Army Research Office under Award number: W911NF-26-1-A043.   
D.D. and P.S. acknowledge funding by the Cluster of Excellence “Advanced Imaging of Matter” of the Deutsche Forschungsgemeinschaft (DFG) - EXC 2056 - project ID 390715994. The numerical computations were performed using the PHYSnet computational cluster at the University of Hamburg. D.D. gratefully acknowledges the technical support of Martin Stieben.

\appendix

\section{Role of the external trap}\label{app:impurity:freespace}

To complement our investigation of the impact of the impurity on the droplet hosts, we compute the ground-states of the three-component system within the eGPE framework of Equation~(\ref{3comp_eGPEs}) in a box potential, i.e., assuming $\omega=0$. 
For convenience, we restrict our analysis to the eGPE model, as the absence of a harmonic trap leads to droplets with large spatial extent, making full many-body simulations significantly more computationally demanding.

\begin{figure}
\centering
\includegraphics[width=\linewidth]{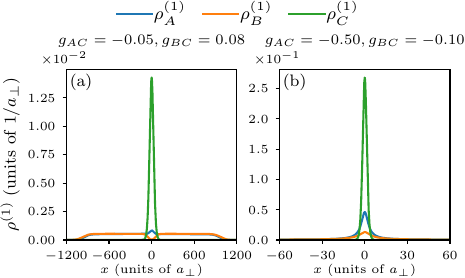}
\caption{Ground{$-$}state density profiles of the different species, $\rho^{(1)}_{\sigma}(x)$, within the eGPE approach in the absence of the harmonic trap, $\omega=0$. The impurity species $\sigma=C$ is (a) weakly and (b) strongly coupled (see legends), with the majority species $\sigma=A,B$. 
The hosts ($\sigma=A,B$) configure in flat{$-$}top structures featuring a hump (dip) for attractive (repulsive) impurity{$-$}droplet couplings. 
The impact of the impurity is more prominent as compared to the trapped setting (see also Figure~\ref{fig5-assymetric-coupling-plus-eGPE-comp}a,c). Other system parameters are the same as in Figure~\ref{fig:fig1-decoupled-impurity}.}
\label{fig:figAppendixGS}
\end{figure}

Paradigmatic ground-state density profiles of the different species are shown in \mbox{Figure~\ref{fig:figAppendixGS}} for different mixed impurity--droplet couplings, corresponding to the weak (Figure~\ref{fig:figAppendixGS}a) and relatively strong (Figure~\ref{fig:figAppendixGS}b) interaction cases. 
A few noticeable differences can be discerned with respect to the trapped scenario (see also Figure~\ref{fig5-assymetric-coupling-plus-eGPE-comp}a,c).
First, the overall shape of the droplet hosts has a clear elongated flat-top shape (Figure~\ref{fig:figAppendixGS}), which is naturally attributed to the flat geometry. 
This behavior is in sharp contrast to the parabolic form profile occurring in the presence of the harmonic trap (Figure~\ref{fig5-assymetric-coupling-plus-eGPE-comp}), which accordingly favors densities with larger peak. 
The latter arguably renders the influence of the impurity to the droplet majority components less prominent in the harmonic trap case.  
For this reason, the imprint of the impurity on the droplet background is evident also for weak interactions, as it can be readily deduced from the densities depicted in Figure~\ref{fig:figAppendixGS}a within the box trap. 
Moreover, as in the main text, attractive interactions between the impurity and one of the majority components facilitate a local accumulation of atoms in the vicinity of the impurity, while the remaining atoms of that component populate the flat-top tails.
As a consequence, the majority density exhibits a hump at the position of the impurity, which is enhanced for stronger attractions (see Figure~\ref{fig:figAppendixGS}b). 
On the other hand, for repulsive couplings, the distribution of the majority component features a density dip in the neighborhood of the impurity, reflecting that majority species atoms are expelled from the impurity.

\section{Impurity dressing}\label{app:dressing}

The impurity atom immersed in the many-body bath is expected to be dressed by the excitations of the majority components for finite interactions between the impurity and at least one of the majority species. 
It is known that such a process can lead to the concept of quasi-particles~\cite{Landau1933}, such as polarons, which have been observed and characterized using ultracold Bose gases (see, e.g., the reviews of Refs.~\cite{scazza2022repulsive,bose-gases-low-d-mistakidis,Grusdt_2025,Tajima}).
However, the notion of quasi-particles is largely unexplored in droplet-bearing environments, and so far they have not been observed in this context.  

Our three-component setting provides a natural starting point toward this direction. 
A common measure to estimate the dressing of the impurity is the wave function overlap between the non-interacting and the interacting state of the impurity atom~\cite{Grusdt_2025}. 
For simplicity, in what follows, we consider the case of a symmetrically coupled impurity to the majority components, namely $g_{AC}=g_{BC} \equiv g$, as discussed in Section~\ref{sec:Symmetrical}. 
Accordingly, the fidelity reads  
\begin{equation} 
F(g)= |\langle \Psi_C(g=0)|\Psi_C(g) \rangle|^2, 
\end{equation}
where $|\Psi_C(g)\rangle$ denotes the wave function of the impurity at a given finite coupling strength, and $|\Psi(g=0) \rangle$ represents the corresponding wave function in the decoupled limit. 
This measure is reminiscent of the contrast of a single impurity, which has been studied extensively for repulsive Bose gases~\cite{bose-gases-low-d-mistakidis,Grusdt_2025,Knap} and it is known to be related to the so-called Ramsey response~\cite{Cetina}.
Our computations rely on both the eGPE framework, which accounts for correlations to first order via the LHY term, and the standard mean-field approach, where all correlations are ignored. 

\begin{figure}
\centering
\includegraphics[width=\linewidth]{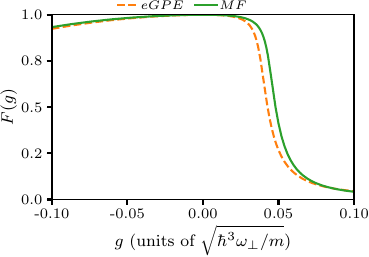}
\caption{Fidelity, $F(g)$, of the impurity's ground{$-$}state wave function between the decoupled ($g=0$) and symmetrically coupled impurity to the majority components. The fidelity is calculated within the eGPE and the mean{$-$}field approximations (see legend) in the presence of a harmonic trap. 
It is evident that $F(g)$ deviates from unity for $g \neq 0$, manifesting the impurity's dressing by the majority components. A more prominent decay takes place for $g>0$ due to the phase separation of the impurity with the majority species. 
The remaining system parameters are the same with the ones of Figure~\ref{fig:fig1-decoupled-impurity}.}
\label{fig:egpe-fidelity-near-PS}
\end{figure}

The ensuing fidelity within the eGPE method, and for $\omega=0.005$, is illustrated in Figure~\ref{fig:egpe-fidelity-near-PS} with respect to different coupling strengths between the impurity and the bath components. 
Within the eGPE approach, we observe that $F(g)$ deviates from unity independently of the interaction sign. 
This trend is indicative of the underlying dressing mechanism from the excitations of the majority components. 
Additionally, it can be seen that the fidelity drop is more prominent for repulsive interactions as compared to attractive ones. 
For relatively weak repulsions, the impurity's distribution exhibits a gradual broadening (Figure~\ref{fig:fig2-rho_sym_gAB_-0_02}b) and hence $F(g)<1$. 
In fact, for $g>0.035$, the fidelity features a dramatic decay toward zero, which suggests a tendency toward an orthogonality catastrophe event~\cite{mistakidis2020many,Knap,Goold}. 
This behavior is traced back to the phase-separation of the impurity from the bath components, in which the impurity's distribution splits into two symmetric humps lying at the trap edges, and the majority components reside at the trap center (see the corresponding density distributions in Figure~\ref{fig:fig2-rho_sym_gAB_-0_02}b and the density overlap in Figure~\ref{fig:overlap-mf-egpe}). 
As such, the impurity escapes from its hosts and it becomes undressed. 
Turning to attractive interactions, the impurity is positioned around the trap center and features a localization tendency at stronger $g<0$ (Figure~\ref{fig:fig2-rho_sym_gAB_-0_02}b). 
This justifies the observed fidelity alterations, signifying the dressing of the impurity for all considered attractive interactions.

The fidelity behavior obtained using the mean-field approximation is somewhat altered as compared to its eGPE counterpart described above. 
In particular, the mean-field fidelity shows a quantitatively slower decay compared to the eGPE prediction. 
This can be understood by the fact that in the mean-field approximation the impurity's broadening and localization at repulsive and attractive interactions occur at slightly larger mean-field couplings due to the absence of the attractive LHY contribution. 
The same holds for the phase-separation process (see also Figure~\ref{fig:overlap-mf-egpe}), and hence the tendency of $F(g>0) \to 0$ is less pronounced at the mean-field level.  

\bibliography{literature.bib}

\begin{thebibliography}{92}%
\makeatletter
\providecommand \@ifxundefined [1]{%
 \@ifx{#1\undefined}
}%
\providecommand \@ifnum [1]{%
 \ifnum #1\expandafter \@firstoftwo
 \else \expandafter \@secondoftwo
 \fi
}%
\providecommand \@ifx [1]{%
 \ifx #1\expandafter \@firstoftwo
 \else \expandafter \@secondoftwo
 \fi
}%
\providecommand \natexlab [1]{#1}%
\providecommand \enquote  [1]{``#1''}%
\providecommand \bibnamefont  [1]{#1}%
\providecommand \bibfnamefont [1]{#1}%
\providecommand \citenamefont [1]{#1}%
\providecommand \href@noop [0]{\@secondoftwo}%
\providecommand \href [0]{\begingroup \@sanitize@url \@href}%
\providecommand \@href[1]{\@@startlink{#1}\@@href}%
\providecommand \@@href[1]{\endgroup#1\@@endlink}%
\providecommand \@sanitize@url [0]{\catcode `\\12\catcode `\$12\catcode `\&12\catcode `\#12\catcode `\^12\catcode `\_12\catcode `\%12\relax}%
\providecommand \@@startlink[1]{}%
\providecommand \@@endlink[0]{}%
\providecommand \url  [0]{\begingroup\@sanitize@url \@url }%
\providecommand \@url [1]{\endgroup\@href {#1}{\urlprefix }}%
\providecommand \urlprefix  [0]{URL }%
\providecommand \Eprint [0]{\href }%
\providecommand \doibase [0]{https://doi.org/}%
\providecommand \selectlanguage [0]{\@gobble}%
\providecommand \bibinfo  [0]{\@secondoftwo}%
\providecommand \bibfield  [0]{\@secondoftwo}%
\providecommand \translation [1]{[#1]}%
\providecommand \BibitemOpen [0]{}%
\providecommand \bibitemStop [0]{}%
\providecommand \bibitemNoStop [0]{.\EOS\space}%
\providecommand \EOS [0]{\spacefactor3000\relax}%
\providecommand \BibitemShut  [1]{\csname bibitem#1\endcsname}%
\let\auto@bib@innerbib\@empty
\bibitem [{\citenamefont {Luo}\ \emph {et~al.}(2021)\citenamefont {Luo}, \citenamefont {Pang}, \citenamefont {Liu}, \citenamefont {Li},\ and\ \citenamefont {Malomed}}]{luo2021new}%
  \BibitemOpen
  \bibfield  {author} {\bibinfo {author} {\bibfnamefont {Z.-H.}\ \bibnamefont {Luo}}, \bibinfo {author} {\bibfnamefont {W.}~\bibnamefont {Pang}}, \bibinfo {author} {\bibfnamefont {B.}~\bibnamefont {Liu}}, \bibinfo {author} {\bibfnamefont {Y.-Y.}\ \bibnamefont {Li}},\ and\ \bibinfo {author} {\bibfnamefont {B.~A.}\ \bibnamefont {Malomed}},\ }\bibfield  {title} {\bibinfo {title} {A new form of liquid matter: Quantum droplets},\ }\href {https://doi.org/10.1007/s11467-020-1020-2} {\bibfield  {journal} {\bibinfo  {journal} {Front. Phys.}\ }\textbf {\bibinfo {volume} {16}},\ \bibinfo {pages} {1} (\bibinfo {year} {2021})}\BibitemShut {NoStop}%
\bibitem [{\citenamefont {B{\"o}ttcher}\ \emph {et~al.}(2020)\citenamefont {B{\"o}ttcher}, \citenamefont {Schmidt}, \citenamefont {Hertkorn}, \citenamefont {N~g}, \citenamefont {Graham}, \citenamefont {Guo}, \citenamefont {Langen},\ and\ \citenamefont {Pfau}}]{bottcher2020new}%
  \BibitemOpen
  \bibfield  {author} {\bibinfo {author} {\bibfnamefont {F.}~\bibnamefont {B{\"o}ttcher}}, \bibinfo {author} {\bibfnamefont {J.-N.}\ \bibnamefont {Schmidt}}, \bibinfo {author} {\bibfnamefont {J.}~\bibnamefont {Hertkorn}}, \bibinfo {author} {\bibfnamefont {K.~S.~H.}\ \bibnamefont {N~g}}, \bibinfo {author} {\bibfnamefont {S.~D.}\ \bibnamefont {Graham}}, \bibinfo {author} {\bibfnamefont {M.}~\bibnamefont {Guo}}, \bibinfo {author} {\bibfnamefont {T.}~\bibnamefont {Langen}},\ and\ \bibinfo {author} {\bibfnamefont {T.}~\bibnamefont {Pfau}},\ }\bibfield  {title} {\bibinfo {title} {New states of matter with fine-tuned interactions: Quantum droplets and dipolar supersolids},\ }\href {https://doi.org/10.1088/1361-6633/abc9ab} {\bibfield  {journal} {\bibinfo  {journal} {Rep. Progr. Phys.}\ }\textbf {\bibinfo {volume} {84}},\ \bibinfo {pages} {012403} (\bibinfo {year} {2020})}\BibitemShut {NoStop}%
\bibitem [{\citenamefont {Mistakidis}\ \emph {et~al.}(2023)\citenamefont {Mistakidis}, \citenamefont {Volosniev}, \citenamefont {Barfknecht}, \citenamefont {Fogarty}, \citenamefont {Busch}, \citenamefont {Foerster}, \citenamefont {Schmelcher},\ and\ \citenamefont {Zinner}}]{bose-gases-low-d-mistakidis}%
  \BibitemOpen
  \bibfield  {author} {\bibinfo {author} {\bibfnamefont {S.}~\bibnamefont {Mistakidis}}, \bibinfo {author} {\bibfnamefont {A.}~\bibnamefont {Volosniev}}, \bibinfo {author} {\bibfnamefont {R.}~\bibnamefont {Barfknecht}}, \bibinfo {author} {\bibfnamefont {T.}~\bibnamefont {Fogarty}}, \bibinfo {author} {\bibfnamefont {T.}~\bibnamefont {Busch}}, \bibinfo {author} {\bibfnamefont {A.}~\bibnamefont {Foerster}}, \bibinfo {author} {\bibfnamefont {P.}~\bibnamefont {Schmelcher}},\ and\ \bibinfo {author} {\bibfnamefont {N.}~\bibnamefont {Zinner}},\ }\bibfield  {title} {\bibinfo {title} {Few-body {Bose} gases in low dimensions—a laboratory for quantum dynamics},\ }\href {https://doi.org/https://doi.org/10.1016/j.physrep.2023.10.004} {\bibfield  {journal} {\bibinfo  {journal} {Phys. Rep.}\ }\textbf {\bibinfo {volume} {1042}},\ \bibinfo {pages} {1} (\bibinfo {year} {2023})}\BibitemShut {NoStop}%
\bibitem [{\citenamefont {Petrov}(2015)}]{petrov-droplets-2015-PhysRevLett.115.155302}%
  \BibitemOpen
  \bibfield  {author} {\bibinfo {author} {\bibfnamefont {D.~S.}\ \bibnamefont {Petrov}},\ }\bibfield  {title} {\bibinfo {title} {Quantum mechanical stabilization of a collapsing {Bose-Bose} mixture},\ }\href {https://doi.org/10.1103/PhysRevLett.115.155302} {\bibfield  {journal} {\bibinfo  {journal} {Phys. Rev. Lett.}\ }\textbf {\bibinfo {volume} {115}},\ \bibinfo {pages} {155302} (\bibinfo {year} {2015})}\BibitemShut {NoStop}%
\bibitem [{\citenamefont {Petrov}\ and\ \citenamefont {Astrakharchik}(2016)}]{petrov-ultra-dilute-2016-PhysRevLett.117.100401}%
  \BibitemOpen
  \bibfield  {author} {\bibinfo {author} {\bibfnamefont {D.~S.}\ \bibnamefont {Petrov}}\ and\ \bibinfo {author} {\bibfnamefont {G.~E.}\ \bibnamefont {Astrakharchik}},\ }\bibfield  {title} {\bibinfo {title} {Ultradilute low-dimensional liquids},\ }\href {https://doi.org/10.1103/PhysRevLett.117.100401} {\bibfield  {journal} {\bibinfo  {journal} {Phys. Rev. Lett.}\ }\textbf {\bibinfo {volume} {117}},\ \bibinfo {pages} {100401} (\bibinfo {year} {2016})}\BibitemShut {NoStop}%
\bibitem [{\citenamefont {Ferrier-Barbut}\ \emph {et~al.}(2016)\citenamefont {Ferrier-Barbut}, \citenamefont {Kadau}, \citenamefont {Schmitt}, \citenamefont {Wenzel},\ and\ \citenamefont {Pfau}}]{ferrier2016observation}%
  \BibitemOpen
  \bibfield  {author} {\bibinfo {author} {\bibfnamefont {I.}~\bibnamefont {Ferrier-Barbut}}, \bibinfo {author} {\bibfnamefont {H.}~\bibnamefont {Kadau}}, \bibinfo {author} {\bibfnamefont {M.}~\bibnamefont {Schmitt}}, \bibinfo {author} {\bibfnamefont {M.}~\bibnamefont {Wenzel}},\ and\ \bibinfo {author} {\bibfnamefont {T.}~\bibnamefont {Pfau}},\ }\bibfield  {title} {\bibinfo {title} {Observation of quantum droplets in a strongly dipolar {B}ose gas},\ }\href {https://doi.org/10.1103/PhysRevLett.116.215301} {\bibfield  {journal} {\bibinfo  {journal} {Phys. Rev. Lett.}\ }\textbf {\bibinfo {volume} {116}},\ \bibinfo {pages} {215301} (\bibinfo {year} {2016})}\BibitemShut {NoStop}%
\bibitem [{\citenamefont {Chomaz}\ \emph {et~al.}(2022)\citenamefont {Chomaz}, \citenamefont {Ferrier-Barbut}, \citenamefont {Ferlaino}, \citenamefont {Laburthe-Tolra}, \citenamefont {Lev},\ and\ \citenamefont {Pfau}}]{chomaz2022dipolar}%
  \BibitemOpen
  \bibfield  {author} {\bibinfo {author} {\bibfnamefont {L.}~\bibnamefont {Chomaz}}, \bibinfo {author} {\bibfnamefont {I.}~\bibnamefont {Ferrier-Barbut}}, \bibinfo {author} {\bibfnamefont {F.}~\bibnamefont {Ferlaino}}, \bibinfo {author} {\bibfnamefont {B.}~\bibnamefont {Laburthe-Tolra}}, \bibinfo {author} {\bibfnamefont {B.~L.}\ \bibnamefont {Lev}},\ and\ \bibinfo {author} {\bibfnamefont {T.}~\bibnamefont {Pfau}},\ }\bibfield  {title} {\bibinfo {title} {Dipolar physics: A review of experiments with magnetic quantum gases},\ }\href {https://doi.org/10.1088/1361-6633/aca814} {\bibfield  {journal} {\bibinfo  {journal} {Rep. Progr. Phys.}\ }\textbf {\bibinfo {volume} {86}},\ \bibinfo {pages} {026401} (\bibinfo {year} {2022})}\BibitemShut {NoStop}%
\bibitem [{\citenamefont {Cabrera}\ \emph {et~al.}(2018)\citenamefont {Cabrera}, \citenamefont {Tanzi}, \citenamefont {Sanz}, \citenamefont {Naylor}, \citenamefont {Thomas}, \citenamefont {Cheiney},\ and\ \citenamefont {Tarruell}}]{CabreraTarruellDropExp}%
  \BibitemOpen
  \bibfield  {author} {\bibinfo {author} {\bibfnamefont {C.~R.}\ \bibnamefont {Cabrera}}, \bibinfo {author} {\bibfnamefont {L.}~\bibnamefont {Tanzi}}, \bibinfo {author} {\bibfnamefont {J.}~\bibnamefont {Sanz}}, \bibinfo {author} {\bibfnamefont {B.}~\bibnamefont {Naylor}}, \bibinfo {author} {\bibfnamefont {P.}~\bibnamefont {Thomas}}, \bibinfo {author} {\bibfnamefont {P.}~\bibnamefont {Cheiney}},\ and\ \bibinfo {author} {\bibfnamefont {L.}~\bibnamefont {Tarruell}},\ }\bibfield  {title} {\bibinfo {title} {Quantum liquid droplets in a mixture of {Bose-Einstein} condensates},\ }\href {https://doi.org/10.1126/science.aao5686} {\bibfield  {journal} {\bibinfo  {journal} {Science}\ }\textbf {\bibinfo {volume} {359}},\ \bibinfo {pages} {301} (\bibinfo {year} {2018})}\BibitemShut {NoStop}%
\bibitem [{\citenamefont {Cheiney}\ \emph {et~al.}(2018)\citenamefont {Cheiney}, \citenamefont {Cabrera}, \citenamefont {Sanz}, \citenamefont {Naylor}, \citenamefont {Tanzi},\ and\ \citenamefont {Tarruell}}]{CheineyTarruellDropExp}%
  \BibitemOpen
  \bibfield  {author} {\bibinfo {author} {\bibfnamefont {P.}~\bibnamefont {Cheiney}}, \bibinfo {author} {\bibfnamefont {C.~R.}\ \bibnamefont {Cabrera}}, \bibinfo {author} {\bibfnamefont {J.}~\bibnamefont {Sanz}}, \bibinfo {author} {\bibfnamefont {B.}~\bibnamefont {Naylor}}, \bibinfo {author} {\bibfnamefont {L.}~\bibnamefont {Tanzi}},\ and\ \bibinfo {author} {\bibfnamefont {L.}~\bibnamefont {Tarruell}},\ }\bibfield  {title} {\bibinfo {title} {Bright soliton to quantum droplet transition in a mixture of {Bose-Einstein} condensates},\ }\href {https://doi.org/10.1103/PhysRevLett.120.135301} {\bibfield  {journal} {\bibinfo  {journal} {Phys. Rev. Lett.}\ }\textbf {\bibinfo {volume} {120}},\ \bibinfo {pages} {135301} (\bibinfo {year} {2018})}\BibitemShut {NoStop}%
\bibitem [{\citenamefont {Semeghini}\ \emph {et~al.}(2018)\citenamefont {Semeghini}, \citenamefont {Ferioli}, \citenamefont {Masi}, \citenamefont {Mazzinghi}, \citenamefont {Wolswijk}, \citenamefont {Minardi}, \citenamefont {Modugno}, \citenamefont {Modugno}, \citenamefont {Inguscio},\ and\ \citenamefont {Fattori}}]{SemeghiniFattoriDropExp}%
  \BibitemOpen
  \bibfield  {author} {\bibinfo {author} {\bibfnamefont {G.}~\bibnamefont {Semeghini}}, \bibinfo {author} {\bibfnamefont {G.}~\bibnamefont {Ferioli}}, \bibinfo {author} {\bibfnamefont {L.}~\bibnamefont {Masi}}, \bibinfo {author} {\bibfnamefont {C.}~\bibnamefont {Mazzinghi}}, \bibinfo {author} {\bibfnamefont {L.}~\bibnamefont {Wolswijk}}, \bibinfo {author} {\bibfnamefont {F.}~\bibnamefont {Minardi}}, \bibinfo {author} {\bibfnamefont {M.}~\bibnamefont {Modugno}}, \bibinfo {author} {\bibfnamefont {G.}~\bibnamefont {Modugno}}, \bibinfo {author} {\bibfnamefont {M.}~\bibnamefont {Inguscio}},\ and\ \bibinfo {author} {\bibfnamefont {M.}~\bibnamefont {Fattori}},\ }\bibfield  {title} {\bibinfo {title} {Self-bound quantum droplets of atomic mixtures in free space},\ }\href {https://doi.org/10.1103/PhysRevLett.120.235301} {\bibfield  {journal} {\bibinfo  {journal} {Phys. Rev. Lett.}\ }\textbf {\bibinfo {volume} {120}},\ \bibinfo {pages} {235301} (\bibinfo {year} {2018})}\BibitemShut {NoStop}%
\bibitem [{\citenamefont {Guo}\ \emph {et~al.}(2021)\citenamefont {Guo}, \citenamefont {Jia}, \citenamefont {Li}, \citenamefont {Ma}, \citenamefont {Hutson}, \citenamefont {Cui},\ and\ \citenamefont {Wang}}]{GuoHeteroDrop2021}%
  \BibitemOpen
  \bibfield  {author} {\bibinfo {author} {\bibfnamefont {Z.}~\bibnamefont {Guo}}, \bibinfo {author} {\bibfnamefont {F.}~\bibnamefont {Jia}}, \bibinfo {author} {\bibfnamefont {L.}~\bibnamefont {Li}}, \bibinfo {author} {\bibfnamefont {Y.}~\bibnamefont {Ma}}, \bibinfo {author} {\bibfnamefont {J.~M.}\ \bibnamefont {Hutson}}, \bibinfo {author} {\bibfnamefont {X.}~\bibnamefont {Cui}},\ and\ \bibinfo {author} {\bibfnamefont {D.}~\bibnamefont {Wang}},\ }\bibfield  {title} {\bibinfo {title} {{Lee-Huang-Yang} effects in the ultracold mixture of $^{23}\mathrm{Na}$ and $^{87}\mathrm{Rb}$ with attractive interspecies interactions},\ }\href {https://doi.org/10.1103/PhysRevResearch.3.033247} {\bibfield  {journal} {\bibinfo  {journal} {Phys. Rev. Res.}\ }\textbf {\bibinfo {volume} {3}},\ \bibinfo {pages} {033247} (\bibinfo {year} {2021})}\BibitemShut {NoStop}%
\bibitem [{\citenamefont {D'Errico}\ \emph {et~al.}(2019)\citenamefont {D'Errico}, \citenamefont {Burchianti}, \citenamefont {Prevedelli}, \citenamefont {Salasnich}, \citenamefont {Ancilotto}, \citenamefont {Modugno}, \citenamefont {Minardi},\ and\ \citenamefont {Fort}}]{FortHeteroExp}%
  \BibitemOpen
  \bibfield  {author} {\bibinfo {author} {\bibfnamefont {C.}~\bibnamefont {D'Errico}}, \bibinfo {author} {\bibfnamefont {A.}~\bibnamefont {Burchianti}}, \bibinfo {author} {\bibfnamefont {M.}~\bibnamefont {Prevedelli}}, \bibinfo {author} {\bibfnamefont {L.}~\bibnamefont {Salasnich}}, \bibinfo {author} {\bibfnamefont {F.}~\bibnamefont {Ancilotto}}, \bibinfo {author} {\bibfnamefont {M.}~\bibnamefont {Modugno}}, \bibinfo {author} {\bibfnamefont {F.}~\bibnamefont {Minardi}},\ and\ \bibinfo {author} {\bibfnamefont {C.}~\bibnamefont {Fort}},\ }\bibfield  {title} {\bibinfo {title} {Observation of quantum droplets in a heteronuclear bosonic mixture},\ }\href {https://doi.org/10.1103/PhysRevResearch.1.033155} {\bibfield  {journal} {\bibinfo  {journal} {Phys. Rev. Res.}\ }\textbf {\bibinfo {volume} {1}},\ \bibinfo {pages} {033155} (\bibinfo {year} {2019})}\BibitemShut {NoStop}%
\bibitem [{\citenamefont {Cavicchioli}\ \emph {et~al.}(2025)\citenamefont {Cavicchioli}, \citenamefont {Fort}, \citenamefont {Ancilotto}, \citenamefont {Modugno}, \citenamefont {Minardi},\ and\ \citenamefont {Burchianti}}]{Cavicchioli}%
  \BibitemOpen
  \bibfield  {author} {\bibinfo {author} {\bibfnamefont {L.}~\bibnamefont {Cavicchioli}}, \bibinfo {author} {\bibfnamefont {C.}~\bibnamefont {Fort}}, \bibinfo {author} {\bibfnamefont {F.}~\bibnamefont {Ancilotto}}, \bibinfo {author} {\bibfnamefont {M.}~\bibnamefont {Modugno}}, \bibinfo {author} {\bibfnamefont {F.}~\bibnamefont {Minardi}},\ and\ \bibinfo {author} {\bibfnamefont {A.}~\bibnamefont {Burchianti}},\ }\bibfield  {title} {\bibinfo {title} {Dynamical formation of multiple quantum droplets in a {Bose-Bose} mixture},\ }\href {https://doi.org/10.1103/PhysRevLett.134.093401} {\bibfield  {journal} {\bibinfo  {journal} {Phys. Rev. Lett.}\ }\textbf {\bibinfo {volume} {134}},\ \bibinfo {pages} {093401} (\bibinfo {year} {2025})}\BibitemShut {NoStop}%
\bibitem [{\citenamefont {Lee}\ \emph {et~al.}(1957)\citenamefont {Lee}, \citenamefont {Huang},\ and\ \citenamefont {Yang}}]{LeeHuangYang1957}%
  \BibitemOpen
  \bibfield  {author} {\bibinfo {author} {\bibfnamefont {T.~D.}\ \bibnamefont {Lee}}, \bibinfo {author} {\bibfnamefont {K.}~\bibnamefont {Huang}},\ and\ \bibinfo {author} {\bibfnamefont {C.~N.}\ \bibnamefont {Yang}},\ }\bibfield  {title} {\bibinfo {title} {Eigenvalues and eigenfunctions of a {Bose} system of hard spheres and its low-temperature properties},\ }\href {https://doi.org/10.1103/PhysRev.106.1135} {\bibfield  {journal} {\bibinfo  {journal} {Phys. Rev.}\ }\textbf {\bibinfo {volume} {106}},\ \bibinfo {pages} {1135} (\bibinfo {year} {1957})}\BibitemShut {NoStop}%
\bibitem [{\citenamefont {Zin}\ \emph {et~al.}(2018)\citenamefont {Zin}, \citenamefont {Pylak}, \citenamefont {Wasak}, \citenamefont {Gajda},\ and\ \citenamefont {Idziaszek}}]{dim_crossover_Zin}%
  \BibitemOpen
  \bibfield  {author} {\bibinfo {author} {\bibfnamefont {P.}~\bibnamefont {Zin}}, \bibinfo {author} {\bibfnamefont {M.}~\bibnamefont {Pylak}}, \bibinfo {author} {\bibfnamefont {T.}~\bibnamefont {Wasak}}, \bibinfo {author} {\bibfnamefont {M.}~\bibnamefont {Gajda}},\ and\ \bibinfo {author} {\bibfnamefont {Z.}~\bibnamefont {Idziaszek}},\ }\bibfield  {title} {\bibinfo {title} {Quantum {B}ose-{B}ose droplets at a dimensional crossover},\ }\href {https://doi.org/10.1103/PhysRevA.98.051603} {\bibfield  {journal} {\bibinfo  {journal} {Phys. Rev. A}\ }\textbf {\bibinfo {volume} {98}},\ \bibinfo {pages} {051603} (\bibinfo {year} {2018})}\BibitemShut {NoStop}%
\bibitem [{\citenamefont {Ilg}\ \emph {et~al.}(2018)\citenamefont {Ilg}, \citenamefont {Kumlin}, \citenamefont {Santos}, \citenamefont {Petrov},\ and\ \citenamefont {B\"uchler}}]{Ilg_crossover_2018}%
  \BibitemOpen
  \bibfield  {author} {\bibinfo {author} {\bibfnamefont {T.}~\bibnamefont {Ilg}}, \bibinfo {author} {\bibfnamefont {J.}~\bibnamefont {Kumlin}}, \bibinfo {author} {\bibfnamefont {L.}~\bibnamefont {Santos}}, \bibinfo {author} {\bibfnamefont {D.~S.}\ \bibnamefont {Petrov}},\ and\ \bibinfo {author} {\bibfnamefont {H.~P.}\ \bibnamefont {B\"uchler}},\ }\bibfield  {title} {\bibinfo {title} {Dimensional crossover for the beyond-mean-field correction in {B}ose gases},\ }\href {https://doi.org/10.1103/PhysRevA.98.051604} {\bibfield  {journal} {\bibinfo  {journal} {Phys. Rev. A}\ }\textbf {\bibinfo {volume} {98}},\ \bibinfo {pages} {051604} (\bibinfo {year} {2018})}\BibitemShut {NoStop}%
\bibitem [{\citenamefont {Pelayo}\ \emph {et~al.}(2025{\natexlab{a}})\citenamefont {Pelayo}, \citenamefont {Bougas}, \citenamefont {Fogarty}, \citenamefont {Busch},\ and\ \citenamefont {Mistakidis}}]{Pelayo_crossover}%
  \BibitemOpen
  \bibfield  {author} {\bibinfo {author} {\bibfnamefont {J.~C.}\ \bibnamefont {Pelayo}}, \bibinfo {author} {\bibfnamefont {G.}~\bibnamefont {Bougas}}, \bibinfo {author} {\bibfnamefont {T.}~\bibnamefont {Fogarty}}, \bibinfo {author} {\bibfnamefont {T.}~\bibnamefont {Busch}},\ and\ \bibinfo {author} {\bibfnamefont {S.~I.}\ \bibnamefont {Mistakidis}},\ }\bibfield  {title} {\bibinfo {title} {{Phases and dynamics of quantum droplets in the crossover to two-dimensions}},\ }\href {https://doi.org/10.21468/SciPostPhys.18.4.129} {\bibfield  {journal} {\bibinfo  {journal} {SciPost Phys.}\ }\textbf {\bibinfo {volume} {18}},\ \bibinfo {pages} {129} (\bibinfo {year} {2025}{\natexlab{a}})}\BibitemShut {NoStop}%
\bibitem [{\citenamefont {Zhang}\ \emph {et~al.}(2025)\citenamefont {Zhang}, \citenamefont {Ye}, \citenamefont {Zhou},\ and\ \citenamefont {Liang}}]{zhang2025self}%
  \BibitemOpen
  \bibfield  {author} {\bibinfo {author} {\bibfnamefont {Y.}~\bibnamefont {Zhang}}, \bibinfo {author} {\bibfnamefont {X.}~\bibnamefont {Ye}}, \bibinfo {author} {\bibfnamefont {Z.}~\bibnamefont {Zhou}},\ and\ \bibinfo {author} {\bibfnamefont {Z.}~\bibnamefont {Liang}},\ }\bibfield  {title} {\bibinfo {title} {Self-consistent effective field theory to nonuniversal {Lee-Huang-Yang} term in quantum droplets},\ }\href@noop {} {\bibfield  {journal} {\bibinfo  {journal} {arXiv:2512.11513}\ } (\bibinfo {year} {2025})}\BibitemShut {NoStop}%
\bibitem [{\citenamefont {Englezos}\ \emph {et~al.}(2025)\citenamefont {Englezos}, \citenamefont {Schmelcher},\ and\ \citenamefont {Mistakidis}}]{ilias-simos-SciPostPhys.19.5.133}%
  \BibitemOpen
  \bibfield  {author} {\bibinfo {author} {\bibfnamefont {I.~A.}\ \bibnamefont {Englezos}}, \bibinfo {author} {\bibfnamefont {P.}~\bibnamefont {Schmelcher}},\ and\ \bibinfo {author} {\bibfnamefont {S.~I.}\ \bibnamefont {Mistakidis}},\ }\bibfield  {title} {\bibinfo {title} {{Multicomponent one-dimensional quantum droplets across the mean-field stability regime}},\ }\href {https://doi.org/10.21468/SciPostPhys.19.5.133} {\bibfield  {journal} {\bibinfo  {journal} {SciPost Phys.}\ }\textbf {\bibinfo {volume} {19}},\ \bibinfo {pages} {133} (\bibinfo {year} {2025})}\BibitemShut {NoStop}%
\bibitem [{\citenamefont {Astrakharchik}\ and\ \citenamefont {Malomed}(2018)}]{1Ddrops_stat_dyn}%
  \BibitemOpen
  \bibfield  {author} {\bibinfo {author} {\bibfnamefont {G.~E.}\ \bibnamefont {Astrakharchik}}\ and\ \bibinfo {author} {\bibfnamefont {B.~A.}\ \bibnamefont {Malomed}},\ }\bibfield  {title} {\bibinfo {title} {Dynamics of one-dimensional quantum droplets},\ }\href {https://doi.org/10.1103/PhysRevA.98.013631} {\bibfield  {journal} {\bibinfo  {journal} {Phys. Rev. A}\ }\textbf {\bibinfo {volume} {98}},\ \bibinfo {pages} {013631} (\bibinfo {year} {2018})}\BibitemShut {NoStop}%
\bibitem [{\citenamefont {Englezos}\ \emph {et~al.}(2024)\citenamefont {Englezos}, \citenamefont {Schmelcher},\ and\ \citenamefont {Mistakidis}}]{ilias-simos-particle-imbalance-PhysRevA.110.023324}%
  \BibitemOpen
  \bibfield  {author} {\bibinfo {author} {\bibfnamefont {I.~A.}\ \bibnamefont {Englezos}}, \bibinfo {author} {\bibfnamefont {P.}~\bibnamefont {Schmelcher}},\ and\ \bibinfo {author} {\bibfnamefont {S.~I.}\ \bibnamefont {Mistakidis}},\ }\bibfield  {title} {\bibinfo {title} {Particle-imbalanced weakly interacting quantum droplets in one dimension},\ }\href {https://doi.org/10.1103/PhysRevA.110.023324} {\bibfield  {journal} {\bibinfo  {journal} {Phys. Rev. A}\ }\textbf {\bibinfo {volume} {110}},\ \bibinfo {pages} {023324} (\bibinfo {year} {2024})}\BibitemShut {NoStop}%
\bibitem [{\citenamefont {Kartashov}\ and\ \citenamefont {Zezyulin}(2024)}]{Kartashov_multipoles}%
  \BibitemOpen
  \bibfield  {author} {\bibinfo {author} {\bibfnamefont {Y.~V.}\ \bibnamefont {Kartashov}}\ and\ \bibinfo {author} {\bibfnamefont {D.~A.}\ \bibnamefont {Zezyulin}},\ }\bibfield  {title} {\bibinfo {title} {Multipole quantum droplets in quasi-one-dimensional asymmetric mixtures},\ }\href {https://doi.org/10.1103/PhysRevA.110.L021304} {\bibfield  {journal} {\bibinfo  {journal} {Phys. Rev. A}\ }\textbf {\bibinfo {volume} {110}},\ \bibinfo {pages} {L021304} (\bibinfo {year} {2024})}\BibitemShut {NoStop}%
\bibitem [{\citenamefont {Xiao}\ \emph {et~al.}(2026)\citenamefont {Xiao}, \citenamefont {Zhang}, \citenamefont {Liu}, \citenamefont {Du}, \citenamefont {Chen},\ and\ \citenamefont {Zhang}}]{xiao2026one}%
  \BibitemOpen
  \bibfield  {author} {\bibinfo {author} {\bibfnamefont {H.}~\bibnamefont {Xiao}}, \bibinfo {author} {\bibfnamefont {X.}~\bibnamefont {Zhang}}, \bibinfo {author} {\bibfnamefont {J.}~\bibnamefont {Liu}}, \bibinfo {author} {\bibfnamefont {X.}~\bibnamefont {Du}}, \bibinfo {author} {\bibfnamefont {X.-L.}\ \bibnamefont {Chen}},\ and\ \bibinfo {author} {\bibfnamefont {Y.}~\bibnamefont {Zhang}},\ }\bibfield  {title} {\bibinfo {title} {One-dimensional asymmetrically interacting quantum droplets in {Bose-Bose} mixtures},\ }\href@noop {} {\bibfield  {journal} {\bibinfo  {journal} {arXiv:2601.16808}\ } (\bibinfo {year} {2026})}\BibitemShut {NoStop}%
\bibitem [{\citenamefont {Flynn}\ \emph {et~al.}(2023)\citenamefont {Flynn}, \citenamefont {Parisi}, \citenamefont {Billam},\ and\ \citenamefont {Parker}}]{flynn2023quantum}%
  \BibitemOpen
  \bibfield  {author} {\bibinfo {author} {\bibfnamefont {T.~A.}\ \bibnamefont {Flynn}}, \bibinfo {author} {\bibfnamefont {L.}~\bibnamefont {Parisi}}, \bibinfo {author} {\bibfnamefont {T.~P.}\ \bibnamefont {Billam}},\ and\ \bibinfo {author} {\bibfnamefont {N.~G.}\ \bibnamefont {Parker}},\ }\bibfield  {title} {\bibinfo {title} {Quantum droplets in imbalanced atomic mixtures},\ }\href {https://doi.org/10.1103/PhysRevResearch.5.033167} {\bibfield  {journal} {\bibinfo  {journal} {Phys. Rev. Res.}\ }\textbf {\bibinfo {volume} {5}},\ \bibinfo {pages} {033167} (\bibinfo {year} {2023})}\BibitemShut {NoStop}%
\bibitem [{\citenamefont {Flynn}\ \emph {et~al.}(2024)\citenamefont {Flynn}, \citenamefont {Keepfer}, \citenamefont {Parker},\ and\ \citenamefont {Billam}}]{flynn2024harmonically}%
  \BibitemOpen
  \bibfield  {author} {\bibinfo {author} {\bibfnamefont {T.}~\bibnamefont {Flynn}}, \bibinfo {author} {\bibfnamefont {N.}~\bibnamefont {Keepfer}}, \bibinfo {author} {\bibfnamefont {N.}~\bibnamefont {Parker}},\ and\ \bibinfo {author} {\bibfnamefont {T.}~\bibnamefont {Billam}},\ }\bibfield  {title} {\bibinfo {title} {Harmonically trapped imbalanced quantum droplets},\ }\href {https://doi.org/10.1103/PhysRevResearch.6.013209} {\bibfield  {journal} {\bibinfo  {journal} {Phys. Rev. Res.}\ }\textbf {\bibinfo {volume} {6}},\ \bibinfo {pages} {013209} (\bibinfo {year} {2024})}\BibitemShut {NoStop}%
\bibitem [{\citenamefont {Pelayo}\ \emph {et~al.}(2025{\natexlab{b}})\citenamefont {Pelayo}, \citenamefont {Bougas}, \citenamefont {Fogarty}, \citenamefont {Busch},\ and\ \citenamefont {Mistakidis}}]{Pelayo_2025}%
  \BibitemOpen
  \bibfield  {author} {\bibinfo {author} {\bibfnamefont {J.~C.}\ \bibnamefont {Pelayo}}, \bibinfo {author} {\bibfnamefont {G.~A.}\ \bibnamefont {Bougas}}, \bibinfo {author} {\bibfnamefont {T.}~\bibnamefont {Fogarty}}, \bibinfo {author} {\bibfnamefont {T.}~\bibnamefont {Busch}},\ and\ \bibinfo {author} {\bibfnamefont {S.~I.}\ \bibnamefont {Mistakidis}},\ }\bibfield  {title} {\bibinfo {title} {Droplet-gas phases and their dynamical formation in particle imbalanced mixtures},\ }\href {https://doi.org/10.1088/2058-9565/ae1162} {\bibfield  {journal} {\bibinfo  {journal} {Quant. Sci. Techn.}\ }\textbf {\bibinfo {volume} {10}},\ \bibinfo {pages} {045074} (\bibinfo {year} {2025}{\natexlab{b}})}\BibitemShut {NoStop}%
\bibitem [{\citenamefont {Tylutki}\ \emph {et~al.}(2020)\citenamefont {Tylutki}, \citenamefont {Astrakharchik}, \citenamefont {Malomed},\ and\ \citenamefont {Petrov}}]{spectrum1D}%
  \BibitemOpen
  \bibfield  {author} {\bibinfo {author} {\bibfnamefont {M.}~\bibnamefont {Tylutki}}, \bibinfo {author} {\bibfnamefont {G.~E.}\ \bibnamefont {Astrakharchik}}, \bibinfo {author} {\bibfnamefont {B.~A.}\ \bibnamefont {Malomed}},\ and\ \bibinfo {author} {\bibfnamefont {D.~S.}\ \bibnamefont {Petrov}},\ }\bibfield  {title} {\bibinfo {title} {Collective excitations of a one-dimensional quantum droplet},\ }\href {https://doi.org/10.1103/PhysRevA.101.051601} {\bibfield  {journal} {\bibinfo  {journal} {Phys. Rev. A}\ }\textbf {\bibinfo {volume} {101}},\ \bibinfo {pages} {051601} (\bibinfo {year} {2020})}\BibitemShut {NoStop}%
\bibitem [{\citenamefont {Katsimiga}\ \emph {et~al.}(2023{\natexlab{a}})\citenamefont {Katsimiga}, \citenamefont {Mistakidis}, \citenamefont {Koutsokostas}, \citenamefont {Frantzeskakis}, \citenamefont {Carretero-Gonz\'alez},\ and\ \citenamefont {Kevrekidis}}]{Katsimiga_sol_drops}%
  \BibitemOpen
  \bibfield  {author} {\bibinfo {author} {\bibfnamefont {G.~C.}\ \bibnamefont {Katsimiga}}, \bibinfo {author} {\bibfnamefont {S.~I.}\ \bibnamefont {Mistakidis}}, \bibinfo {author} {\bibfnamefont {G.~N.}\ \bibnamefont {Koutsokostas}}, \bibinfo {author} {\bibfnamefont {D.~J.}\ \bibnamefont {Frantzeskakis}}, \bibinfo {author} {\bibfnamefont {R.}~\bibnamefont {Carretero-Gonz\'alez}},\ and\ \bibinfo {author} {\bibfnamefont {P.~G.}\ \bibnamefont {Kevrekidis}},\ }\bibfield  {title} {\bibinfo {title} {Solitary waves in a quantum droplet-bearing system},\ }\href {https://doi.org/10.1103/PhysRevA.107.063308} {\bibfield  {journal} {\bibinfo  {journal} {Phys. Rev. A}\ }\textbf {\bibinfo {volume} {107}},\ \bibinfo {pages} {063308} (\bibinfo {year} {2023}{\natexlab{a}})}\BibitemShut {NoStop}%
\bibitem [{\citenamefont {Charalampidis}\ and\ \citenamefont {Mistakidis}(2025)}]{charalampidis2024two}%
  \BibitemOpen
  \bibfield  {author} {\bibinfo {author} {\bibfnamefont {E.~G.}\ \bibnamefont {Charalampidis}}\ and\ \bibinfo {author} {\bibfnamefont {S.~I.}\ \bibnamefont {Mistakidis}},\ }\bibfield  {title} {\bibinfo {title} {Two-component droplet phases and their spectral stability in one dimension},\ }\href {https://doi.org/10.1103/PhysRevA.111.013318} {\bibfield  {journal} {\bibinfo  {journal} {Phys. Rev. A}\ }\textbf {\bibinfo {volume} {111}},\ \bibinfo {pages} {013318} (\bibinfo {year} {2025})}\BibitemShut {NoStop}%
\bibitem [{\citenamefont {Li}\ \emph {et~al.}(2018)\citenamefont {Li}, \citenamefont {Chen}, \citenamefont {Luo}, \citenamefont {Huang}, \citenamefont {Tan}, \citenamefont {Pang},\ and\ \citenamefont {Malomed}}]{Li_vortex}%
  \BibitemOpen
  \bibfield  {author} {\bibinfo {author} {\bibfnamefont {Y.}~\bibnamefont {Li}}, \bibinfo {author} {\bibfnamefont {Z.}~\bibnamefont {Chen}}, \bibinfo {author} {\bibfnamefont {Z.}~\bibnamefont {Luo}}, \bibinfo {author} {\bibfnamefont {C.}~\bibnamefont {Huang}}, \bibinfo {author} {\bibfnamefont {H.}~\bibnamefont {Tan}}, \bibinfo {author} {\bibfnamefont {W.}~\bibnamefont {Pang}},\ and\ \bibinfo {author} {\bibfnamefont {B.~A.}\ \bibnamefont {Malomed}},\ }\bibfield  {title} {\bibinfo {title} {Two-dimensional vortex quantum droplets},\ }\href {https://doi.org/10.1103/PhysRevA.98.063602} {\bibfield  {journal} {\bibinfo  {journal} {Phys. Rev. A}\ }\textbf {\bibinfo {volume} {98}},\ \bibinfo {pages} {063602} (\bibinfo {year} {2018})}\BibitemShut {NoStop}%
\bibitem [{\citenamefont {Yo\ifmmode~\breve{g}\else \u{g}\fi{}urt}\ \emph {et~al.}(2023)\citenamefont {Yo\ifmmode~\breve{g}\else \u{g}\fi{}urt}, \citenamefont {Tanyeri}, \citenamefont {Kele\ifmmode~\mbox{\c{s}}\else \c{s}\fi{}},\ and\ \citenamefont {Oktel}}]{Yoifmmode}%
  \BibitemOpen
  \bibfield  {author} {\bibinfo {author} {\bibfnamefont {T.~A.}\ \bibnamefont {Yo\ifmmode~\breve{g}\else \u{g}\fi{}urt}}, \bibinfo {author} {\bibfnamefont {U.}~\bibnamefont {Tanyeri}}, \bibinfo {author} {\bibfnamefont {A.}~\bibnamefont {Kele\ifmmode~\mbox{\c{s}}\else \c{s}\fi{}}},\ and\ \bibinfo {author} {\bibfnamefont {M.~O.}\ \bibnamefont {Oktel}},\ }\bibfield  {title} {\bibinfo {title} {Vortex lattices in strongly confined quantum droplets},\ }\href {https://doi.org/10.1103/PhysRevA.108.033315} {\bibfield  {journal} {\bibinfo  {journal} {Phys. Rev. A}\ }\textbf {\bibinfo {volume} {108}},\ \bibinfo {pages} {033315} (\bibinfo {year} {2023})}\BibitemShut {NoStop}%
\bibitem [{\citenamefont {Tengstrand}\ \emph {et~al.}(2019)\citenamefont {Tengstrand}, \citenamefont {St\"urmer}, \citenamefont {Karabulut},\ and\ \citenamefont {Reimann}}]{Tengstrand_vortex}%
  \BibitemOpen
  \bibfield  {author} {\bibinfo {author} {\bibfnamefont {M.~N.}\ \bibnamefont {Tengstrand}}, \bibinfo {author} {\bibfnamefont {P.}~\bibnamefont {St\"urmer}}, \bibinfo {author} {\bibfnamefont {E.~O.}\ \bibnamefont {Karabulut}},\ and\ \bibinfo {author} {\bibfnamefont {S.~M.}\ \bibnamefont {Reimann}},\ }\bibfield  {title} {\bibinfo {title} {Rotating binary {Bose-Einstein} condensates and vortex clusters in quantum droplets},\ }\href {https://doi.org/10.1103/PhysRevLett.123.160405} {\bibfield  {journal} {\bibinfo  {journal} {Phys. Rev. Lett.}\ }\textbf {\bibinfo {volume} {123}},\ \bibinfo {pages} {160405} (\bibinfo {year} {2019})}\BibitemShut {NoStop}%
\bibitem [{\citenamefont {Bougas}\ \emph {et~al.}(2024)\citenamefont {Bougas}, \citenamefont {Katsimiga}, \citenamefont {Kevrekidis},\ and\ \citenamefont {Mistakidis}}]{Bougas_vortex}%
  \BibitemOpen
  \bibfield  {author} {\bibinfo {author} {\bibfnamefont {G.}~\bibnamefont {Bougas}}, \bibinfo {author} {\bibfnamefont {G.~C.}\ \bibnamefont {Katsimiga}}, \bibinfo {author} {\bibfnamefont {P.~G.}\ \bibnamefont {Kevrekidis}},\ and\ \bibinfo {author} {\bibfnamefont {S.~I.}\ \bibnamefont {Mistakidis}},\ }\bibfield  {title} {\bibinfo {title} {Stability and dynamics of nonlinear excitations in a two-dimensional droplet-bearing environment},\ }\href {https://doi.org/10.1103/PhysRevA.110.033317} {\bibfield  {journal} {\bibinfo  {journal} {Phys. Rev. A}\ }\textbf {\bibinfo {volume} {110}},\ \bibinfo {pages} {033317} (\bibinfo {year} {2024})}\BibitemShut {NoStop}%
\bibitem [{\citenamefont {Chandramouli}\ \emph {et~al.}(2024)\citenamefont {Chandramouli}, \citenamefont {Mistakidis}, \citenamefont {Katsimiga},\ and\ \citenamefont {Kevrekidis}}]{Chandramouli_shocks}%
  \BibitemOpen
  \bibfield  {author} {\bibinfo {author} {\bibfnamefont {S.}~\bibnamefont {Chandramouli}}, \bibinfo {author} {\bibfnamefont {S.~I.}\ \bibnamefont {Mistakidis}}, \bibinfo {author} {\bibfnamefont {G.~C.}\ \bibnamefont {Katsimiga}},\ and\ \bibinfo {author} {\bibfnamefont {P.~G.}\ \bibnamefont {Kevrekidis}},\ }\bibfield  {title} {\bibinfo {title} {Dispersive shock waves in a one-dimensional droplet-bearing environment},\ }\href {https://doi.org/10.1103/PhysRevA.110.023304} {\bibfield  {journal} {\bibinfo  {journal} {Phys. Rev. A}\ }\textbf {\bibinfo {volume} {110}},\ \bibinfo {pages} {023304} (\bibinfo {year} {2024})}\BibitemShut {NoStop}%
\bibitem [{\citenamefont {Chandramouli}\ \emph {et~al.}(2026)\citenamefont {Chandramouli}, \citenamefont {Mistakidis}, \citenamefont {Katsimiga}, \citenamefont {Ratliff}, \citenamefont {Frantzeskakis},\ and\ \citenamefont {Kevrekidis}}]{Chandramouli_RWs}%
  \BibitemOpen
  \bibfield  {author} {\bibinfo {author} {\bibfnamefont {S.}~\bibnamefont {Chandramouli}}, \bibinfo {author} {\bibfnamefont {S.~I.}\ \bibnamefont {Mistakidis}}, \bibinfo {author} {\bibfnamefont {G.~C.}\ \bibnamefont {Katsimiga}}, \bibinfo {author} {\bibfnamefont {D.~J.}\ \bibnamefont {Ratliff}}, \bibinfo {author} {\bibfnamefont {D.~J.}\ \bibnamefont {Frantzeskakis}},\ and\ \bibinfo {author} {\bibfnamefont {P.~G.}\ \bibnamefont {Kevrekidis}},\ }\bibfield  {title} {\bibinfo {title} {Rogue waves in extended {Gross-Pitaevskii} models with a {Lee-Huang-Yang} correction},\ }\href {https://doi.org/10.1103/7rlg-z74h} {\bibfield  {journal} {\bibinfo  {journal} {Phys. Rev. A}\ }\textbf {\bibinfo {volume} {113}},\ \bibinfo {pages} {013308} (\bibinfo {year} {2026})}\BibitemShut {NoStop}%
\bibitem [{\citenamefont {Mithun}\ \emph {et~al.}(2020)\citenamefont {Mithun}, \citenamefont {Maluckov}, \citenamefont {Kasamatsu}, \citenamefont {Malomed},\ and\ \citenamefont {Khare}}]{mithun2020modulational}%
  \BibitemOpen
  \bibfield  {author} {\bibinfo {author} {\bibfnamefont {T.}~\bibnamefont {Mithun}}, \bibinfo {author} {\bibfnamefont {A.}~\bibnamefont {Maluckov}}, \bibinfo {author} {\bibfnamefont {K.}~\bibnamefont {Kasamatsu}}, \bibinfo {author} {\bibfnamefont {B.~A.}\ \bibnamefont {Malomed}},\ and\ \bibinfo {author} {\bibfnamefont {A.}~\bibnamefont {Khare}},\ }\bibfield  {title} {\bibinfo {title} {Modulational instability, inter-component asymmetry, and formation of quantum droplets in one-dimensional binary {Bose} gases},\ }\href@noop {} {\bibfield  {journal} {\bibinfo  {journal} {Symmetry}\ }\textbf {\bibinfo {volume} {12}},\ \bibinfo {pages} {174} (\bibinfo {year} {2020})}\BibitemShut {NoStop}%
\bibitem [{\citenamefont {Otajonov}\ \emph {et~al.}(2022)\citenamefont {Otajonov}, \citenamefont {Tsoy},\ and\ \citenamefont {Abdullaev}}]{Otajonov_MI_drops}%
  \BibitemOpen
  \bibfield  {author} {\bibinfo {author} {\bibfnamefont {S.~R.}\ \bibnamefont {Otajonov}}, \bibinfo {author} {\bibfnamefont {E.~N.}\ \bibnamefont {Tsoy}},\ and\ \bibinfo {author} {\bibfnamefont {F.~K.}\ \bibnamefont {Abdullaev}},\ }\bibfield  {title} {\bibinfo {title} {Modulational instability and quantum droplets in a two-dimensional {Bose-Einstein} condensate},\ }\href {https://doi.org/10.1103/PhysRevA.106.033309} {\bibfield  {journal} {\bibinfo  {journal} {Phys. Rev. A}\ }\textbf {\bibinfo {volume} {106}},\ \bibinfo {pages} {033309} (\bibinfo {year} {2022})}\BibitemShut {NoStop}%
\bibitem [{\citenamefont {Kartashov}\ \emph {et~al.}(2022)\citenamefont {Kartashov}, \citenamefont {Lashkin}, \citenamefont {Modugno},\ and\ \citenamefont {Torner}}]{kartashov2022spinor}%
  \BibitemOpen
  \bibfield  {author} {\bibinfo {author} {\bibfnamefont {Y.~V.}\ \bibnamefont {Kartashov}}, \bibinfo {author} {\bibfnamefont {V.~M.}\ \bibnamefont {Lashkin}}, \bibinfo {author} {\bibfnamefont {M.}~\bibnamefont {Modugno}},\ and\ \bibinfo {author} {\bibfnamefont {L.}~\bibnamefont {Torner}},\ }\bibfield  {title} {\bibinfo {title} {Spinor-induced instability of kinks, holes and quantum droplets},\ }\href@noop {} {\bibfield  {journal} {\bibinfo  {journal} {New J. Phys.}\ }\textbf {\bibinfo {volume} {24}},\ \bibinfo {pages} {073012} (\bibinfo {year} {2022})}\BibitemShut {NoStop}%
\bibitem [{\citenamefont {Katsimiga}\ \emph {et~al.}(2023{\natexlab{b}})\citenamefont {Katsimiga}, \citenamefont {Mistakidis}, \citenamefont {Malomed}, \citenamefont {Frantzeskakis}, \citenamefont {Carretero-Gonzalez},\ and\ \citenamefont {Kevrekidis}}]{katsimiga2023interactions}%
  \BibitemOpen
  \bibfield  {author} {\bibinfo {author} {\bibfnamefont {G.~C.}\ \bibnamefont {Katsimiga}}, \bibinfo {author} {\bibfnamefont {S.~I.}\ \bibnamefont {Mistakidis}}, \bibinfo {author} {\bibfnamefont {B.~A.}\ \bibnamefont {Malomed}}, \bibinfo {author} {\bibfnamefont {D.~J.}\ \bibnamefont {Frantzeskakis}}, \bibinfo {author} {\bibfnamefont {R.}~\bibnamefont {Carretero-Gonzalez}},\ and\ \bibinfo {author} {\bibfnamefont {P.~G.}\ \bibnamefont {Kevrekidis}},\ }\bibfield  {title} {\bibinfo {title} {Interactions and dynamics of one-dimensional droplets, bubbles and kinks},\ }\href@noop {} {\bibfield  {journal} {\bibinfo  {journal} {Condensed Matter}\ }\textbf {\bibinfo {volume} {8}},\ \bibinfo {pages} {67} (\bibinfo {year} {2023}{\natexlab{b}})}\BibitemShut {NoStop}%
\bibitem [{\citenamefont {Mistakidis}\ \emph {et~al.}(2025)\citenamefont {Mistakidis}, \citenamefont {Bougas}, \citenamefont {Katsimiga},\ and\ \citenamefont {Kevrekidis}}]{Mistakidis_kink}%
  \BibitemOpen
  \bibfield  {author} {\bibinfo {author} {\bibfnamefont {S.~I.}\ \bibnamefont {Mistakidis}}, \bibinfo {author} {\bibfnamefont {G.}~\bibnamefont {Bougas}}, \bibinfo {author} {\bibfnamefont {G.~C.}\ \bibnamefont {Katsimiga}},\ and\ \bibinfo {author} {\bibfnamefont {P.~G.}\ \bibnamefont {Kevrekidis}},\ }\bibfield  {title} {\bibinfo {title} {Generic transverse stability of kink structures in atomic and optical nonlinear media with competing attractive and repulsive interactions},\ }\href {https://doi.org/10.1103/PhysRevLett.134.123402} {\bibfield  {journal} {\bibinfo  {journal} {Phys. Rev. Lett.}\ }\textbf {\bibinfo {volume} {134}},\ \bibinfo {pages} {123402} (\bibinfo {year} {2025})}\BibitemShut {NoStop}%
\bibitem [{\citenamefont {Parisi}\ and\ \citenamefont {Giorgini}(2020)}]{ParisiGiorginiMonteCarlo}%
  \BibitemOpen
  \bibfield  {author} {\bibinfo {author} {\bibfnamefont {L.}~\bibnamefont {Parisi}}\ and\ \bibinfo {author} {\bibfnamefont {S.}~\bibnamefont {Giorgini}},\ }\bibfield  {title} {\bibinfo {title} {Quantum droplets in one-dimensional {Bose} mixtures: A quantum monte carlo study},\ }\href {https://doi.org/10.1103/PhysRevA.102.023318} {\bibfield  {journal} {\bibinfo  {journal} {Phys. Rev. A}\ }\textbf {\bibinfo {volume} {102}},\ \bibinfo {pages} {023318} (\bibinfo {year} {2020})}\BibitemShut {NoStop}%
\bibitem [{\citenamefont {Parisi}\ \emph {et~al.}(2019)\citenamefont {Parisi}, \citenamefont {Astrakharchik},\ and\ \citenamefont {Giorgini}}]{ParisiMonteCarlo2019}%
  \BibitemOpen
  \bibfield  {author} {\bibinfo {author} {\bibfnamefont {L.}~\bibnamefont {Parisi}}, \bibinfo {author} {\bibfnamefont {G.~E.}\ \bibnamefont {Astrakharchik}},\ and\ \bibinfo {author} {\bibfnamefont {S.}~\bibnamefont {Giorgini}},\ }\bibfield  {title} {\bibinfo {title} {Liquid state of one-dimensional {Bose} mixtures: A quantum monte carlo study},\ }\href {https://doi.org/10.1103/PhysRevLett.122.105302} {\bibfield  {journal} {\bibinfo  {journal} {Phys. Rev. Lett.}\ }\textbf {\bibinfo {volume} {122}},\ \bibinfo {pages} {105302} (\bibinfo {year} {2019})}\BibitemShut {NoStop}%
\bibitem [{\citenamefont {Mistakidis}\ \emph {et~al.}(2021)\citenamefont {Mistakidis}, \citenamefont {Mithun}, \citenamefont {Kevrekidis}, \citenamefont {Sadeghpour},\ and\ \citenamefont {Schmelcher}}]{droplet-harmonic_pot-Mistakidis_2021}%
  \BibitemOpen
  \bibfield  {author} {\bibinfo {author} {\bibfnamefont {S.~I.}\ \bibnamefont {Mistakidis}}, \bibinfo {author} {\bibfnamefont {T.}~\bibnamefont {Mithun}}, \bibinfo {author} {\bibfnamefont {P.~G.}\ \bibnamefont {Kevrekidis}}, \bibinfo {author} {\bibfnamefont {H.~R.}\ \bibnamefont {Sadeghpour}},\ and\ \bibinfo {author} {\bibfnamefont {P.}~\bibnamefont {Schmelcher}},\ }\bibfield  {title} {\bibinfo {title} {Formation and quench of homonuclear and heteronuclear quantum droplets in one dimension},\ }\bibfield  {journal} {\bibinfo  {journal} {Phys. Rev. Res.}\ }\textbf {\bibinfo {volume} {3}},\ \href {https://doi.org/10.1103/physrevresearch.3.043128} {10.1103/physrevresearch.3.043128} (\bibinfo {year} {2021})\BibitemShut {NoStop}%
\bibitem [{\citenamefont {Englezos}\ \emph {et~al.}(2023)\citenamefont {Englezos}, \citenamefont {Mistakidis},\ and\ \citenamefont {Schmelcher}}]{ilias-simos-droplet-excitations-PhysRevA.107.023320}%
  \BibitemOpen
  \bibfield  {author} {\bibinfo {author} {\bibfnamefont {I.~A.}\ \bibnamefont {Englezos}}, \bibinfo {author} {\bibfnamefont {S.~I.}\ \bibnamefont {Mistakidis}},\ and\ \bibinfo {author} {\bibfnamefont {P.}~\bibnamefont {Schmelcher}},\ }\bibfield  {title} {\bibinfo {title} {Correlated dynamics of collective droplet excitations in a one-dimensional harmonic trap},\ }\href {https://doi.org/10.1103/PhysRevA.107.023320} {\bibfield  {journal} {\bibinfo  {journal} {Phys. Rev. A}\ }\textbf {\bibinfo {volume} {107}},\ \bibinfo {pages} {023320} (\bibinfo {year} {2023})}\BibitemShut {NoStop}%
\bibitem [{\citenamefont {Hu}\ and\ \citenamefont {Liu}(2020)}]{Hu_theory}%
  \BibitemOpen
  \bibfield  {author} {\bibinfo {author} {\bibfnamefont {H.}~\bibnamefont {Hu}}\ and\ \bibinfo {author} {\bibfnamefont {X.-J.}\ \bibnamefont {Liu}},\ }\bibfield  {title} {\bibinfo {title} {Consistent theory of self-bound quantum droplets with bosonic pairing},\ }\href {https://doi.org/10.1103/PhysRevLett.125.195302} {\bibfield  {journal} {\bibinfo  {journal} {Phys. Rev. Lett.}\ }\textbf {\bibinfo {volume} {125}},\ \bibinfo {pages} {195302} (\bibinfo {year} {2020})}\BibitemShut {NoStop}%
\bibitem [{\citenamefont {Gu}\ and\ \citenamefont {Yin}(2020)}]{Gu_phonon}%
  \BibitemOpen
  \bibfield  {author} {\bibinfo {author} {\bibfnamefont {Q.}~\bibnamefont {Gu}}\ and\ \bibinfo {author} {\bibfnamefont {L.}~\bibnamefont {Yin}},\ }\bibfield  {title} {\bibinfo {title} {Phonon stability and sound velocity of quantum droplets in a boson mixture},\ }\href {https://doi.org/10.1103/PhysRevB.102.220503} {\bibfield  {journal} {\bibinfo  {journal} {Phys. Rev. B}\ }\textbf {\bibinfo {volume} {102}},\ \bibinfo {pages} {220503} (\bibinfo {year} {2020})}\BibitemShut {NoStop}%
\bibitem [{\citenamefont {Pan}\ \emph {et~al.}(2022)\citenamefont {Pan}, \citenamefont {Yi},\ and\ \citenamefont {Shi}}]{Pan_critical}%
  \BibitemOpen
  \bibfield  {author} {\bibinfo {author} {\bibfnamefont {J.}~\bibnamefont {Pan}}, \bibinfo {author} {\bibfnamefont {S.}~\bibnamefont {Yi}},\ and\ \bibinfo {author} {\bibfnamefont {T.}~\bibnamefont {Shi}},\ }\bibfield  {title} {\bibinfo {title} {Quantum phases of self-bound droplets of {Bose-Bose} mixtures},\ }\href {https://doi.org/10.1103/PhysRevResearch.4.043018} {\bibfield  {journal} {\bibinfo  {journal} {Phys. Rev. Res.}\ }\textbf {\bibinfo {volume} {4}},\ \bibinfo {pages} {043018} (\bibinfo {year} {2022})}\BibitemShut {NoStop}%
\bibitem [{\citenamefont {Ota}\ and\ \citenamefont {Astrakharchik}(2020)}]{Ota_beyond_LHY}%
  \BibitemOpen
  \bibfield  {author} {\bibinfo {author} {\bibfnamefont {M.}~\bibnamefont {Ota}}\ and\ \bibinfo {author} {\bibfnamefont {G.~E.}\ \bibnamefont {Astrakharchik}},\ }\bibfield  {title} {\bibinfo {title} {{Beyond Lee-Huang-Yang description of self-bound Bose mixtures}},\ }\href {https://doi.org/10.21468/SciPostPhys.9.2.020} {\bibfield  {journal} {\bibinfo  {journal} {SciPost Phys.}\ }\textbf {\bibinfo {volume} {9}},\ \bibinfo {pages} {020} (\bibinfo {year} {2020})}\BibitemShut {NoStop}%
\bibitem [{\citenamefont {Englezos}\ \emph {et~al.}(2026)\citenamefont {Englezos}, \citenamefont {Charalampidis}, \citenamefont {Schmelcher},\ and\ \citenamefont {Mistakidis}}]{englezos2026stabilitymixedphasesthreecomponent}%
  \BibitemOpen
  \bibfield  {author} {\bibinfo {author} {\bibfnamefont {I.~A.}\ \bibnamefont {Englezos}}, \bibinfo {author} {\bibfnamefont {E.~G.}\ \bibnamefont {Charalampidis}}, \bibinfo {author} {\bibfnamefont {P.}~\bibnamefont {Schmelcher}},\ and\ \bibinfo {author} {\bibfnamefont {S.~I.}\ \bibnamefont {Mistakidis}},\ }\href {https://arxiv.org/abs/2601.04950} {\bibinfo {title} {Stability and mixed phases of three-component droplets in one dimension}} (\bibinfo {year} {2026}),\ \Eprint {https://arxiv.org/abs/2601.04950} {arXiv:2601.04950 [cond-mat.quant-gas]} \BibitemShut {NoStop}%
\bibitem [{\citenamefont {Ma}\ \emph {et~al.}(2021)\citenamefont {Ma}, \citenamefont {Peng},\ and\ \citenamefont {Cui}}]{ma2021borromean}%
  \BibitemOpen
  \bibfield  {author} {\bibinfo {author} {\bibfnamefont {Y.}~\bibnamefont {Ma}}, \bibinfo {author} {\bibfnamefont {C.}~\bibnamefont {Peng}},\ and\ \bibinfo {author} {\bibfnamefont {X.}~\bibnamefont {Cui}},\ }\bibfield  {title} {\bibinfo {title} {Borromean droplet in three-component ultracold {Bose} gases},\ }\href@noop {} {\bibfield  {journal} {\bibinfo  {journal} {Phys. Rev. Lett.}\ }\textbf {\bibinfo {volume} {127}},\ \bibinfo {pages} {043002} (\bibinfo {year} {2021})}\BibitemShut {NoStop}%
\bibitem [{\citenamefont {Ma}\ and\ \citenamefont {Cui}(2025)}]{Cui2025_3CompDroplets}%
  \BibitemOpen
  \bibfield  {author} {\bibinfo {author} {\bibfnamefont {Y.}~\bibnamefont {Ma}}\ and\ \bibinfo {author} {\bibfnamefont {X.}~\bibnamefont {Cui}},\ }\bibfield  {title} {\bibinfo {title} {Shell-shaped quantum droplet in a three-component ultracold {Bose} gas},\ }\href {https://doi.org/10.1103/PhysRevLett.134.043402} {\bibfield  {journal} {\bibinfo  {journal} {Phys. Rev. Lett.}\ }\textbf {\bibinfo {volume} {134}},\ \bibinfo {pages} {043402} (\bibinfo {year} {2025})}\BibitemShut {NoStop}%
\bibitem [{\citenamefont {Edler}\ \emph {et~al.}(2022)\citenamefont {Edler}, \citenamefont {Ardila}, \citenamefont {Cabrera},\ and\ \citenamefont {Santos}}]{Edler_buoyancy}%
  \BibitemOpen
  \bibfield  {author} {\bibinfo {author} {\bibfnamefont {D.}~\bibnamefont {Edler}}, \bibinfo {author} {\bibfnamefont {L.~A. P.~n.}\ \bibnamefont {Ardila}}, \bibinfo {author} {\bibfnamefont {C.~R.}\ \bibnamefont {Cabrera}},\ and\ \bibinfo {author} {\bibfnamefont {L.}~\bibnamefont {Santos}},\ }\bibfield  {title} {\bibinfo {title} {Anomalous buoyancy of quantum bubbles in immiscible bose mixtures},\ }\href {https://doi.org/10.1103/PhysRevResearch.4.033017} {\bibfield  {journal} {\bibinfo  {journal} {Phys. Rev. Res.}\ }\textbf {\bibinfo {volume} {4}},\ \bibinfo {pages} {033017} (\bibinfo {year} {2022})}\BibitemShut {NoStop}%
\bibitem [{\citenamefont {Scazza}\ \emph {et~al.}(2022)\citenamefont {Scazza}, \citenamefont {Zaccanti}, \citenamefont {Massignan}, \citenamefont {Parish},\ and\ \citenamefont {Levinsen}}]{scazza2022repulsive}%
  \BibitemOpen
  \bibfield  {author} {\bibinfo {author} {\bibfnamefont {F.}~\bibnamefont {Scazza}}, \bibinfo {author} {\bibfnamefont {M.}~\bibnamefont {Zaccanti}}, \bibinfo {author} {\bibfnamefont {P.}~\bibnamefont {Massignan}}, \bibinfo {author} {\bibfnamefont {M.~M.}\ \bibnamefont {Parish}},\ and\ \bibinfo {author} {\bibfnamefont {J.}~\bibnamefont {Levinsen}},\ }\bibfield  {title} {\bibinfo {title} {Repulsive {Fermi} and {Bose} polarons in quantum gases},\ }\href@noop {} {\bibfield  {journal} {\bibinfo  {journal} {Atoms}\ }\textbf {\bibinfo {volume} {10}},\ \bibinfo {pages} {55} (\bibinfo {year} {2022})}\BibitemShut {NoStop}%
\bibitem [{\citenamefont {Grusdt}\ \emph {et~al.}(2025)\citenamefont {Grusdt}, \citenamefont {Mostaan}, \citenamefont {Demler},\ and\ \citenamefont {Ardila}}]{Grusdt_2025}%
  \BibitemOpen
  \bibfield  {author} {\bibinfo {author} {\bibfnamefont {F.}~\bibnamefont {Grusdt}}, \bibinfo {author} {\bibfnamefont {N.}~\bibnamefont {Mostaan}}, \bibinfo {author} {\bibfnamefont {E.}~\bibnamefont {Demler}},\ and\ \bibinfo {author} {\bibfnamefont {L.~A.~P.}\ \bibnamefont {Ardila}},\ }\bibfield  {title} {\bibinfo {title} {Impurities and polarons in bosonic quantum gases: A review on recent progress},\ }\href {https://doi.org/10.1088/1361-6633/add94b} {\bibfield  {journal} {\bibinfo  {journal} {Rep. Progr. Phys.}\ }\textbf {\bibinfo {volume} {88}},\ \bibinfo {pages} {066401} (\bibinfo {year} {2025})}\BibitemShut {NoStop}%
\bibitem [{\citenamefont {Sinha}\ \emph {et~al.}(2023)\citenamefont {Sinha}, \citenamefont {Biswas}, \citenamefont {Santos},\ and\ \citenamefont {Sinha}}]{sinha-imp-1d-droplet}%
  \BibitemOpen
  \bibfield  {author} {\bibinfo {author} {\bibfnamefont {S.}~\bibnamefont {Sinha}}, \bibinfo {author} {\bibfnamefont {S.}~\bibnamefont {Biswas}}, \bibinfo {author} {\bibfnamefont {L.}~\bibnamefont {Santos}},\ and\ \bibinfo {author} {\bibfnamefont {S.}~\bibnamefont {Sinha}},\ }\bibfield  {title} {\bibinfo {title} {Impurities in quasi-one-dimensional droplets of binary {Bose} mixtures},\ }\href {https://doi.org/10.1103/PhysRevA.108.023311} {\bibfield  {journal} {\bibinfo  {journal} {Phys. Rev. A}\ }\textbf {\bibinfo {volume} {108}},\ \bibinfo {pages} {023311} (\bibinfo {year} {2023})}\BibitemShut {NoStop}%
\bibitem [{\citenamefont {Abdullaev}\ and\ \citenamefont {Galimzyanov}(2020)}]{abdullaev2020bosonic}%
  \BibitemOpen
  \bibfield  {author} {\bibinfo {author} {\bibfnamefont {F.~K.}\ \bibnamefont {Abdullaev}}\ and\ \bibinfo {author} {\bibfnamefont {R.}~\bibnamefont {Galimzyanov}},\ }\bibfield  {title} {\bibinfo {title} {Bosonic impurity in a one-dimensional quantum droplet in the {Bose--Bose} mixture},\ }\href {https://iopscience.iop.org/article/10.1088/1361-6455/ab9659/meta?casa_token=dQ7pKuoxb0IAAAAA:20DA_plbpOHD9RCI2X8X4-jXNkqruXAKwf-kbhO2DFSWkb1Gv4xtGQbHjBmLoZdPQ0KN6yLRDt2bXjg2mqyweCka1VLW} {\bibfield  {journal} {\bibinfo  {journal} {J. Phys. B: At., Mol. Opt. Phys.}\ }\textbf {\bibinfo {volume} {53}},\ \bibinfo {pages} {165301} (\bibinfo {year} {2020})}\BibitemShut {NoStop}%
\bibitem [{\citenamefont {Bighin}\ \emph {et~al.}(2022)\citenamefont {Bighin}, \citenamefont {Burchianti}, \citenamefont {Minardi},\ and\ \citenamefont {Macr\`{\i}}}]{Bighin_localization}%
  \BibitemOpen
  \bibfield  {author} {\bibinfo {author} {\bibfnamefont {G.}~\bibnamefont {Bighin}}, \bibinfo {author} {\bibfnamefont {A.}~\bibnamefont {Burchianti}}, \bibinfo {author} {\bibfnamefont {F.}~\bibnamefont {Minardi}},\ and\ \bibinfo {author} {\bibfnamefont {T.}~\bibnamefont {Macr\`{\i}}},\ }\bibfield  {title} {\bibinfo {title} {Impurity in a heteronuclear two-component {Bose} mixture},\ }\href {https://doi.org/10.1103/PhysRevA.106.023301} {\bibfield  {journal} {\bibinfo  {journal} {Phys. Rev. A}\ }\textbf {\bibinfo {volume} {106}},\ \bibinfo {pages} {023301} (\bibinfo {year} {2022})}\BibitemShut {NoStop}%
\bibitem [{\citenamefont {Debnath}\ \emph {et~al.}(2023)\citenamefont {Debnath}, \citenamefont {Khan},\ and\ \citenamefont {Malomed}}]{debnath2023interaction}%
  \BibitemOpen
  \bibfield  {author} {\bibinfo {author} {\bibfnamefont {A.}~\bibnamefont {Debnath}}, \bibinfo {author} {\bibfnamefont {A.}~\bibnamefont {Khan}},\ and\ \bibinfo {author} {\bibfnamefont {B.}~\bibnamefont {Malomed}},\ }\bibfield  {title} {\bibinfo {title} {Interaction of one-dimensional quantum droplets with potential wells and barriers},\ }\href {https://www.sciencedirect.com/science/article/abs/pii/S1007570423003787} {\bibfield  {journal} {\bibinfo  {journal} {Commun. Nonlinear Sci. Numer. Simul.}\ }\textbf {\bibinfo {volume} {126}},\ \bibinfo {pages} {107457} (\bibinfo {year} {2023})}\BibitemShut {NoStop}%
\bibitem [{\citenamefont {Bristy}\ \emph {et~al.}(2025)\citenamefont {Bristy}, \citenamefont {Bougas}, \citenamefont {Katsimiga},\ and\ \citenamefont {Mistakidis}}]{Bristy_defect_drops}%
  \BibitemOpen
  \bibfield  {author} {\bibinfo {author} {\bibfnamefont {F.}~\bibnamefont {Bristy}}, \bibinfo {author} {\bibfnamefont {G.}~\bibnamefont {Bougas}}, \bibinfo {author} {\bibfnamefont {G.}~\bibnamefont {Katsimiga}},\ and\ \bibinfo {author} {\bibfnamefont {S.}~\bibnamefont {Mistakidis}},\ }\bibfield  {title} {\bibinfo {title} {Localization and splitting of a quantum droplet with a potential defect},\ }\href {https://doi.org/https://doi.org/10.1016/j.chaos.2025.117383} {\bibfield  {journal} {\bibinfo  {journal} {Chaos Solitons \& Fractals}\ }\textbf {\bibinfo {volume} {201}},\ \bibinfo {pages} {117383} (\bibinfo {year} {2025})}\BibitemShut {NoStop}%
\bibitem [{\citenamefont {Wenzel}\ \emph {et~al.}(2018)\citenamefont {Wenzel}, \citenamefont {Pfau},\ and\ \citenamefont {Ferrier-Barbut}}]{wenzel2018fermionic}%
  \BibitemOpen
  \bibfield  {author} {\bibinfo {author} {\bibfnamefont {M.}~\bibnamefont {Wenzel}}, \bibinfo {author} {\bibfnamefont {T.}~\bibnamefont {Pfau}},\ and\ \bibinfo {author} {\bibfnamefont {I.}~\bibnamefont {Ferrier-Barbut}},\ }\bibfield  {title} {\bibinfo {title} {A fermionic impurity in a dipolar quantum droplet},\ }\href {https://iopscience.iop.org/article/10.1088/1402-4896/aadd72/meta?casa_token=7NeSBwiKKqEAAAAA:bp1h0KuDXiy0Wg3PVKXIz1OcqXs1IX-UA0TMETJDNbtptMXUkWnPAVhDTMUk--mLYCUrGE-wF2sRay96Dl5-o4DtsaAj} {\bibfield  {journal} {\bibinfo  {journal} {Phys. Scripta}\ }\textbf {\bibinfo {volume} {93}},\ \bibinfo {pages} {104004} (\bibinfo {year} {2018})}\BibitemShut {NoStop}%
\bibitem [{\citenamefont {Pelayo}\ \emph {et~al.}(2024)\citenamefont {Pelayo}, \citenamefont {Fogarty}, \citenamefont {Busch},\ and\ \citenamefont {Mistakidis}}]{Pelayo_ferm_imp}%
  \BibitemOpen
  \bibfield  {author} {\bibinfo {author} {\bibfnamefont {J.~C.}\ \bibnamefont {Pelayo}}, \bibinfo {author} {\bibfnamefont {T.}~\bibnamefont {Fogarty}}, \bibinfo {author} {\bibfnamefont {T.}~\bibnamefont {Busch}},\ and\ \bibinfo {author} {\bibfnamefont {S.~I.}\ \bibnamefont {Mistakidis}},\ }\bibfield  {title} {\bibinfo {title} {Phases and dynamics of few fermionic impurities immersed in two-dimensional boson droplets},\ }\href {https://doi.org/10.1103/PhysRevResearch.6.033219} {\bibfield  {journal} {\bibinfo  {journal} {Phys. Rev. Res.}\ }\textbf {\bibinfo {volume} {6}},\ \bibinfo {pages} {033219} (\bibinfo {year} {2024})}\BibitemShut {NoStop}%
\bibitem [{\citenamefont {Cao}\ \emph {et~al.}(2017)\citenamefont {Cao}, \citenamefont {Bolsinger}, \citenamefont {Mistakidis}, \citenamefont {Koutentakis}, \citenamefont {Kr{\"o}nke}, \citenamefont {Schurer},\ and\ \citenamefont {Schmelcher}}]{cao2017}%
  \BibitemOpen
  \bibfield  {author} {\bibinfo {author} {\bibfnamefont {L.}~\bibnamefont {Cao}}, \bibinfo {author} {\bibfnamefont {V.}~\bibnamefont {Bolsinger}}, \bibinfo {author} {\bibfnamefont {S.~I.}\ \bibnamefont {Mistakidis}}, \bibinfo {author} {\bibfnamefont {G.~M.}\ \bibnamefont {Koutentakis}}, \bibinfo {author} {\bibfnamefont {S.}~\bibnamefont {Kr{\"o}nke}}, \bibinfo {author} {\bibfnamefont {J.~M.}\ \bibnamefont {Schurer}},\ and\ \bibinfo {author} {\bibfnamefont {P.}~\bibnamefont {Schmelcher}},\ }\bibfield  {title} {\bibinfo {title} {A unified ab initio approach to the correlated quantum dynamics of ultracold fermionic and bosonic mixtures},\ }\href {https://doi.org/10.1063/1.4993512} {\bibfield  {journal} {\bibinfo  {journal} {J. Chem. Phys.}\ }\textbf {\bibinfo {volume} {147}},\ \bibinfo {pages} {044106} (\bibinfo {year} {2017})}\BibitemShut {NoStop}%
\bibitem [{\citenamefont {Pethick}\ and\ \citenamefont {Smith}(2008)}]{Pethick_Smith_2008}%
  \BibitemOpen
  \bibfield  {author} {\bibinfo {author} {\bibfnamefont {C.~J.}\ \bibnamefont {Pethick}}\ and\ \bibinfo {author} {\bibfnamefont {H.}~\bibnamefont {Smith}},\ }\href@noop {} {\emph {\bibinfo {title} {Bose–Einstein Condensation in Dilute Gases}}},\ \bibinfo {edition} {2nd}\ ed.\ (\bibinfo  {publisher} {Cambridge University Press},\ \bibinfo {year} {2008})\BibitemShut {NoStop}%
\bibitem [{\citenamefont {G\"orlitz}\ \emph {et~al.}(2001)\citenamefont {G\"orlitz}, \citenamefont {Vogels}, \citenamefont {Leanhardt}, \citenamefont {Raman}, \citenamefont {Gustavson}, \citenamefont {Abo-Shaeer}, \citenamefont {Chikkatur}, \citenamefont {Gupta}, \citenamefont {Inouye}, \citenamefont {Rosenband},\ and\ \citenamefont {Ketterle}}]{Ketterle2001LowDexp}%
  \BibitemOpen
  \bibfield  {author} {\bibinfo {author} {\bibfnamefont {A.}~\bibnamefont {G\"orlitz}}, \bibinfo {author} {\bibfnamefont {J.~M.}\ \bibnamefont {Vogels}}, \bibinfo {author} {\bibfnamefont {A.~E.}\ \bibnamefont {Leanhardt}}, \bibinfo {author} {\bibfnamefont {C.}~\bibnamefont {Raman}}, \bibinfo {author} {\bibfnamefont {T.~L.}\ \bibnamefont {Gustavson}}, \bibinfo {author} {\bibfnamefont {J.~R.}\ \bibnamefont {Abo-Shaeer}}, \bibinfo {author} {\bibfnamefont {A.~P.}\ \bibnamefont {Chikkatur}}, \bibinfo {author} {\bibfnamefont {S.}~\bibnamefont {Gupta}}, \bibinfo {author} {\bibfnamefont {S.}~\bibnamefont {Inouye}}, \bibinfo {author} {\bibfnamefont {T.}~\bibnamefont {Rosenband}},\ and\ \bibinfo {author} {\bibfnamefont {W.}~\bibnamefont {Ketterle}},\ }\bibfield  {title} {\bibinfo {title} {Realization of {Bose-Einstein} condensates in lower dimensions},\ }\href {https://doi.org/10.1103/PhysRevLett.87.130402} {\bibfield  {journal} {\bibinfo  {journal} {Phys. Rev. Lett.}\ }\textbf {\bibinfo {volume} {87}},\ \bibinfo
  {pages} {130402} (\bibinfo {year} {2001})}\BibitemShut {NoStop}%
\bibitem [{\citenamefont {Romero-Ros}\ \emph {et~al.}(2024)\citenamefont {Romero-Ros}, \citenamefont {Katsimiga}, \citenamefont {Mistakidis}, \citenamefont {Mossman}, \citenamefont {Biondini}, \citenamefont {Schmelcher}, \citenamefont {Engels},\ and\ \citenamefont {Kevrekidis}}]{Romero_exp_Peregrine}%
  \BibitemOpen
  \bibfield  {author} {\bibinfo {author} {\bibfnamefont {A.}~\bibnamefont {Romero-Ros}}, \bibinfo {author} {\bibfnamefont {G.~C.}\ \bibnamefont {Katsimiga}}, \bibinfo {author} {\bibfnamefont {S.~I.}\ \bibnamefont {Mistakidis}}, \bibinfo {author} {\bibfnamefont {S.}~\bibnamefont {Mossman}}, \bibinfo {author} {\bibfnamefont {G.}~\bibnamefont {Biondini}}, \bibinfo {author} {\bibfnamefont {P.}~\bibnamefont {Schmelcher}}, \bibinfo {author} {\bibfnamefont {P.}~\bibnamefont {Engels}},\ and\ \bibinfo {author} {\bibfnamefont {P.~G.}\ \bibnamefont {Kevrekidis}},\ }\bibfield  {title} {\bibinfo {title} {Experimental realization of the peregrine soliton in repulsive two-component {Bose-Einstein} condensates},\ }\href {https://doi.org/10.1103/PhysRevLett.132.033402} {\bibfield  {journal} {\bibinfo  {journal} {Phys. Rev. Lett.}\ }\textbf {\bibinfo {volume} {132}},\ \bibinfo {pages} {033402} (\bibinfo {year} {2024})}\BibitemShut {NoStop}%
\bibitem [{\citenamefont {K\"ohler}\ \emph {et~al.}(2006)\citenamefont {K\"ohler}, \citenamefont {G\'oral},\ and\ \citenamefont {Julienne}}]{feshbach1-RevModPhys.78.1311}%
  \BibitemOpen
  \bibfield  {author} {\bibinfo {author} {\bibfnamefont {T.}~\bibnamefont {K\"ohler}}, \bibinfo {author} {\bibfnamefont {K.}~\bibnamefont {G\'oral}},\ and\ \bibinfo {author} {\bibfnamefont {P.~S.}\ \bibnamefont {Julienne}},\ }\bibfield  {title} {\bibinfo {title} {Production of cold molecules via magnetically tunable {Feshbach} resonances},\ }\href {https://doi.org/10.1103/RevModPhys.78.1311} {\bibfield  {journal} {\bibinfo  {journal} {Rev. Mod. Phys.}\ }\textbf {\bibinfo {volume} {78}},\ \bibinfo {pages} {1311} (\bibinfo {year} {2006})}\BibitemShut {NoStop}%
\bibitem [{\citenamefont {Chin}\ \emph {et~al.}(2010)\citenamefont {Chin}, \citenamefont {Grimm}, \citenamefont {Julienne},\ and\ \citenamefont {Tiesinga}}]{feshbach2-RevModPhys.82.1225}%
  \BibitemOpen
  \bibfield  {author} {\bibinfo {author} {\bibfnamefont {C.}~\bibnamefont {Chin}}, \bibinfo {author} {\bibfnamefont {R.}~\bibnamefont {Grimm}}, \bibinfo {author} {\bibfnamefont {P.}~\bibnamefont {Julienne}},\ and\ \bibinfo {author} {\bibfnamefont {E.}~\bibnamefont {Tiesinga}},\ }\bibfield  {title} {\bibinfo {title} {Feshbach resonances in ultracold gases},\ }\href {https://doi.org/10.1103/RevModPhys.82.1225} {\bibfield  {journal} {\bibinfo  {journal} {Rev. Mod. Phys.}\ }\textbf {\bibinfo {volume} {82}},\ \bibinfo {pages} {1225} (\bibinfo {year} {2010})}\BibitemShut {NoStop}%
\bibitem [{\citenamefont {Olshanii}(1998)}]{confinement-resonance-1-PhysRevLett.81.938}%
  \BibitemOpen
  \bibfield  {author} {\bibinfo {author} {\bibfnamefont {M.}~\bibnamefont {Olshanii}},\ }\bibfield  {title} {\bibinfo {title} {Atomic scattering in the presence of an external confinement and a gas of impenetrable bosons},\ }\href {https://doi.org/10.1103/PhysRevLett.81.938} {\bibfield  {journal} {\bibinfo  {journal} {Phys. Rev. Lett.}\ }\textbf {\bibinfo {volume} {81}},\ \bibinfo {pages} {938} (\bibinfo {year} {1998})}\BibitemShut {NoStop}%
\bibitem [{\citenamefont {Kr{\"o}nke}\ \emph {et~al.}(2013)\citenamefont {Kr{\"o}nke}, \citenamefont {Cao}, \citenamefont {Vendrell},\ and\ \citenamefont {Schmelcher}}]{kronke2013}%
  \BibitemOpen
  \bibfield  {author} {\bibinfo {author} {\bibfnamefont {S.}~\bibnamefont {Kr{\"o}nke}}, \bibinfo {author} {\bibfnamefont {L.}~\bibnamefont {Cao}}, \bibinfo {author} {\bibfnamefont {O.}~\bibnamefont {Vendrell}},\ and\ \bibinfo {author} {\bibfnamefont {P.}~\bibnamefont {Schmelcher}},\ }\bibfield  {title} {\bibinfo {title} {Non-equilibrium quantum dynamics of ultra-cold atomic mixtures: The multi-layer multi-configuration time-dependent {{Hartree}} method for bosons},\ }\href {https://doi.org/10.1088/1367-2630/15/6/063018} {\bibfield  {journal} {\bibinfo  {journal} {New J. Phys.}\ }\textbf {\bibinfo {volume} {15}},\ \bibinfo {pages} {063018} (\bibinfo {year} {2013})}\BibitemShut {NoStop}%
\bibitem [{\citenamefont {Lode}\ \emph {et~al.}(2020)\citenamefont {Lode}, \citenamefont {L\'ev\^eque}, \citenamefont {Madsen}, \citenamefont {Streltsov},\ and\ \citenamefont {Alon}}]{mctdh-colloquium-RevModPhys.92.011001}%
  \BibitemOpen
  \bibfield  {author} {\bibinfo {author} {\bibfnamefont {A.~U.~J.}\ \bibnamefont {Lode}}, \bibinfo {author} {\bibfnamefont {C.}~\bibnamefont {L\'ev\^eque}}, \bibinfo {author} {\bibfnamefont {L.~B.}\ \bibnamefont {Madsen}}, \bibinfo {author} {\bibfnamefont {A.~I.}\ \bibnamefont {Streltsov}},\ and\ \bibinfo {author} {\bibfnamefont {O.~E.}\ \bibnamefont {Alon}},\ }\bibfield  {title} {\bibinfo {title} {Colloquium: Multiconfigurational time-dependent hartree approaches for indistinguishable particles},\ }\href {https://doi.org/10.1103/RevModPhys.92.011001} {\bibfield  {journal} {\bibinfo  {journal} {Rev. Mod. Phys.}\ }\textbf {\bibinfo {volume} {92}},\ \bibinfo {pages} {011001} (\bibinfo {year} {2020})}\BibitemShut {NoStop}%
\bibitem [{\citenamefont {Vidal}\ and\ \citenamefont {Werner}(2002)}]{vidal2002}%
  \BibitemOpen
  \bibfield  {author} {\bibinfo {author} {\bibfnamefont {G.}~\bibnamefont {Vidal}}\ and\ \bibinfo {author} {\bibfnamefont {R.~F.}\ \bibnamefont {Werner}},\ }\bibfield  {title} {\bibinfo {title} {Computable measure of entanglement},\ }\href {https://doi.org/10.1103/PhysRevA.65.032314} {\bibfield  {journal} {\bibinfo  {journal} {Phys. Rev. A}\ }\textbf {\bibinfo {volume} {65}},\ \bibinfo {pages} {032314} (\bibinfo {year} {2002})}\BibitemShut {NoStop}%
\bibitem [{\citenamefont {Theel}\ \emph {et~al.}(2024)\citenamefont {Theel}, \citenamefont {Mistakidis},\ and\ \citenamefont {Schmelcher}}]{Theel_crossover}%
  \BibitemOpen
  \bibfield  {author} {\bibinfo {author} {\bibfnamefont {F.}~\bibnamefont {Theel}}, \bibinfo {author} {\bibfnamefont {S.~I.}\ \bibnamefont {Mistakidis}},\ and\ \bibinfo {author} {\bibfnamefont {P.}~\bibnamefont {Schmelcher}},\ }\bibfield  {title} {\bibinfo {title} {{Crossover from attractive to repulsive induced interactions and bound states of two distinguishable Bose polarons}},\ }\href {https://doi.org/10.21468/SciPostPhys.16.1.023} {\bibfield  {journal} {\bibinfo  {journal} {SciPost Phys.}\ }\textbf {\bibinfo {volume} {16}},\ \bibinfo {pages} {023} (\bibinfo {year} {2024})}\BibitemShut {NoStop}%
\bibitem [{\citenamefont {Frenkel}(1934)}]{frenkel1934wave}%
  \BibitemOpen
  \bibfield  {author} {\bibinfo {author} {\bibfnamefont {J.}~\bibnamefont {Frenkel}},\ }\href@noop {} {\bibinfo {title} {Wave mechanics; elementary theory}} (\bibinfo {year} {1934})\BibitemShut {NoStop}%
\bibitem [{\citenamefont {Chomaz}\ \emph {et~al.}(2016)\citenamefont {Chomaz}, \citenamefont {Baier}, \citenamefont {Petter}, \citenamefont {Mark}, \citenamefont {W\"achtler}, \citenamefont {Santos},\ and\ \citenamefont {Ferlaino}}]{Chomaz_QFD}%
  \BibitemOpen
  \bibfield  {author} {\bibinfo {author} {\bibfnamefont {L.}~\bibnamefont {Chomaz}}, \bibinfo {author} {\bibfnamefont {S.}~\bibnamefont {Baier}}, \bibinfo {author} {\bibfnamefont {D.}~\bibnamefont {Petter}}, \bibinfo {author} {\bibfnamefont {M.~J.}\ \bibnamefont {Mark}}, \bibinfo {author} {\bibfnamefont {F.}~\bibnamefont {W\"achtler}}, \bibinfo {author} {\bibfnamefont {L.}~\bibnamefont {Santos}},\ and\ \bibinfo {author} {\bibfnamefont {F.}~\bibnamefont {Ferlaino}},\ }\bibfield  {title} {\bibinfo {title} {Quantum-fluctuation-driven crossover from a dilute bose-einstein condensate to a macrodroplet in a dipolar quantum fluid},\ }\href {https://doi.org/10.1103/PhysRevX.6.041039} {\bibfield  {journal} {\bibinfo  {journal} {Phys. Rev. X}\ }\textbf {\bibinfo {volume} {6}},\ \bibinfo {pages} {041039} (\bibinfo {year} {2016})}\BibitemShut {NoStop}%
\bibitem [{\citenamefont {Pitaevskii}\ and\ \citenamefont {Stringari}(2003)}]{pitaevskii2003bose}%
  \BibitemOpen
  \bibfield  {author} {\bibinfo {author} {\bibfnamefont {L.}~\bibnamefont {Pitaevskii}}\ and\ \bibinfo {author} {\bibfnamefont {S.}~\bibnamefont {Stringari}},\ }\href {https://books.google.de/books?id=rIobbOxC4j4C} {\emph {\bibinfo {title} {Bose-Einstein Condensation}}},\ International Series of Monographs on Physics\ (\bibinfo  {publisher} {Clarendon Press},\ \bibinfo {year} {2003})\BibitemShut {NoStop}%
\bibitem [{\citenamefont {Sakmann}\ \emph {et~al.}(2008)\citenamefont {Sakmann}, \citenamefont {Streltsov}, \citenamefont {Alon},\ and\ \citenamefont {Cederbaum}}]{rdm-sankmann-2008-PhysRevA.78.023615}%
  \BibitemOpen
  \bibfield  {author} {\bibinfo {author} {\bibfnamefont {K.}~\bibnamefont {Sakmann}}, \bibinfo {author} {\bibfnamefont {A.~I.}\ \bibnamefont {Streltsov}}, \bibinfo {author} {\bibfnamefont {O.~E.}\ \bibnamefont {Alon}},\ and\ \bibinfo {author} {\bibfnamefont {L.~S.}\ \bibnamefont {Cederbaum}},\ }\bibfield  {title} {\bibinfo {title} {Reduced density matrices and coherence of trapped interacting bosons},\ }\href {https://doi.org/10.1103/PhysRevA.78.023615} {\bibfield  {journal} {\bibinfo  {journal} {Phys. Rev. A}\ }\textbf {\bibinfo {volume} {78}},\ \bibinfo {pages} {023615} (\bibinfo {year} {2008})}\BibitemShut {NoStop}%
\bibitem [{\citenamefont {Mistakidis}\ \emph {et~al.}(2018)\citenamefont {Mistakidis}, \citenamefont {Katsimiga}, \citenamefont {Kevrekidis},\ and\ \citenamefont {Schmelcher}}]{correl-effects-2species-Mistakidis_2018}%
  \BibitemOpen
  \bibfield  {author} {\bibinfo {author} {\bibfnamefont {S.~I.}\ \bibnamefont {Mistakidis}}, \bibinfo {author} {\bibfnamefont {G.~C.}\ \bibnamefont {Katsimiga}}, \bibinfo {author} {\bibfnamefont {P.~G.}\ \bibnamefont {Kevrekidis}},\ and\ \bibinfo {author} {\bibfnamefont {P.}~\bibnamefont {Schmelcher}},\ }\bibfield  {title} {\bibinfo {title} {Correlation effects in the quench-induced phase separation dynamics of a two species ultracold quantum gas},\ }\href {https://doi.org/10.1088/1367-2630/aabc6a} {\bibfield  {journal} {\bibinfo  {journal} {New J. Phys.}\ }\textbf {\bibinfo {volume} {20}},\ \bibinfo {pages} {043052} (\bibinfo {year} {2018})}\BibitemShut {NoStop}%
\bibitem [{\citenamefont {Naraschewski}\ and\ \citenamefont {Glauber}(1999)}]{rdm-Glauber-1999-PhysRevA.59.4595}%
  \BibitemOpen
  \bibfield  {author} {\bibinfo {author} {\bibfnamefont {M.}~\bibnamefont {Naraschewski}}\ and\ \bibinfo {author} {\bibfnamefont {R.~J.}\ \bibnamefont {Glauber}},\ }\bibfield  {title} {\bibinfo {title} {Spatial coherence and density correlations of trapped {Bose} gases},\ }\href {https://doi.org/10.1103/PhysRevA.59.4595} {\bibfield  {journal} {\bibinfo  {journal} {Phys. Rev. A}\ }\textbf {\bibinfo {volume} {59}},\ \bibinfo {pages} {4595} (\bibinfo {year} {1999})}\BibitemShut {NoStop}%
\bibitem [{\citenamefont {Endres}\ \emph {et~al.}(2013)\citenamefont {Endres}, \citenamefont {Cheneau}, \citenamefont {Fukuhara}, \citenamefont {Weitenberg}, \citenamefont {Schau{\ss}}, \citenamefont {Gross}, \citenamefont {Mazza}, \citenamefont {Ba{\~n}uls}, \citenamefont {Pollet}, \citenamefont {Bloch},\ and\ \citenamefont {Kuhr}}]{coh2-exp3-Endres2013}%
  \BibitemOpen
  \bibfield  {author} {\bibinfo {author} {\bibfnamefont {M.}~\bibnamefont {Endres}}, \bibinfo {author} {\bibfnamefont {M.}~\bibnamefont {Cheneau}}, \bibinfo {author} {\bibfnamefont {T.}~\bibnamefont {Fukuhara}}, \bibinfo {author} {\bibfnamefont {C.}~\bibnamefont {Weitenberg}}, \bibinfo {author} {\bibfnamefont {P.}~\bibnamefont {Schau{\ss}}}, \bibinfo {author} {\bibfnamefont {C.}~\bibnamefont {Gross}}, \bibinfo {author} {\bibfnamefont {L.}~\bibnamefont {Mazza}}, \bibinfo {author} {\bibfnamefont {M.~C.}\ \bibnamefont {Ba{\~n}uls}}, \bibinfo {author} {\bibfnamefont {L.}~\bibnamefont {Pollet}}, \bibinfo {author} {\bibfnamefont {I.}~\bibnamefont {Bloch}},\ and\ \bibinfo {author} {\bibfnamefont {S.}~\bibnamefont {Kuhr}},\ }\bibfield  {title} {\bibinfo {title} {Single-site- and single-atom-resolved measurement of correlation functions},\ }\href {https://doi.org/10.1007/s00340-013-5552-9} {\bibfield  {journal} {\bibinfo  {journal} {Appl. Phys. B}\ }\textbf {\bibinfo {volume} {113}},\ \bibinfo {pages} {27} (\bibinfo
  {year} {2013})}\BibitemShut {NoStop}%
\bibitem [{\citenamefont {Tavares}\ \emph {et~al.}(2017)\citenamefont {Tavares}, \citenamefont {Fritsch}, \citenamefont {Telles}, \citenamefont {Hussein}, \citenamefont {Impens}, \citenamefont {Kaiser},\ and\ \citenamefont {Bagnato}}]{coh2-exp1-Tavares_2017}%
  \BibitemOpen
  \bibfield  {author} {\bibinfo {author} {\bibfnamefont {P.~E.~S.}\ \bibnamefont {Tavares}}, \bibinfo {author} {\bibfnamefont {A.~R.}\ \bibnamefont {Fritsch}}, \bibinfo {author} {\bibfnamefont {G.~D.}\ \bibnamefont {Telles}}, \bibinfo {author} {\bibfnamefont {M.~S.}\ \bibnamefont {Hussein}}, \bibinfo {author} {\bibfnamefont {F.}~\bibnamefont {Impens}}, \bibinfo {author} {\bibfnamefont {R.}~\bibnamefont {Kaiser}},\ and\ \bibinfo {author} {\bibfnamefont {V.~S.}\ \bibnamefont {Bagnato}},\ }\bibfield  {title} {\bibinfo {title} {Matter wave speckle observed in an out-of-equilibrium quantum fluid},\ }\href {https://doi.org/10.1073/pnas.1713693114} {\bibfield  {journal} {\bibinfo  {journal} {Proc. Natl. Acad. Sci.}\ }\textbf {\bibinfo {volume} {114}},\ \bibinfo {pages} {12691} (\bibinfo {year} {2017})}\BibitemShut {NoStop}%
\bibitem [{\citenamefont {Nguyen}\ \emph {et~al.}(2019)\citenamefont {Nguyen}, \citenamefont {Tsatsos}, \citenamefont {Luo}, \citenamefont {Lode}, \citenamefont {Telles}, \citenamefont {Bagnato},\ and\ \citenamefont {Hulet}}]{coh2-exp2-Nguyen_2019}%
  \BibitemOpen
  \bibfield  {author} {\bibinfo {author} {\bibfnamefont {J.~H.~V.}\ \bibnamefont {Nguyen}}, \bibinfo {author} {\bibfnamefont {M.~C.}\ \bibnamefont {Tsatsos}}, \bibinfo {author} {\bibfnamefont {D.}~\bibnamefont {Luo}}, \bibinfo {author} {\bibfnamefont {A.~U.~J.}\ \bibnamefont {Lode}}, \bibinfo {author} {\bibfnamefont {G.~D.}\ \bibnamefont {Telles}}, \bibinfo {author} {\bibfnamefont {V.~S.}\ \bibnamefont {Bagnato}},\ and\ \bibinfo {author} {\bibfnamefont {R.~G.}\ \bibnamefont {Hulet}},\ }\bibfield  {title} {\bibinfo {title} {Parametric excitation of a {Bose-Einstein} condensate: From faraday waves to granulation},\ }\bibfield  {journal} {\bibinfo  {journal} {Phys. Rev. X}\ }\textbf {\bibinfo {volume} {9}},\ \href {https://doi.org/10.1103/physrevx.9.011052} {10.1103/physrevx.9.011052} (\bibinfo {year} {2019})\BibitemShut {NoStop}%
\bibitem [{\citenamefont {Mistakidis}\ \emph {et~al.}(2020)\citenamefont {Mistakidis}, \citenamefont {Koutentakis}, \citenamefont {Katsimiga}, \citenamefont {Busch},\ and\ \citenamefont {Schmelcher}}]{mistakidis2020many}%
  \BibitemOpen
  \bibfield  {author} {\bibinfo {author} {\bibfnamefont {S.~I.}\ \bibnamefont {Mistakidis}}, \bibinfo {author} {\bibfnamefont {G.}~\bibnamefont {Koutentakis}}, \bibinfo {author} {\bibfnamefont {G.}~\bibnamefont {Katsimiga}}, \bibinfo {author} {\bibfnamefont {T.}~\bibnamefont {Busch}},\ and\ \bibinfo {author} {\bibfnamefont {P.}~\bibnamefont {Schmelcher}},\ }\bibfield  {title} {\bibinfo {title} {Many-body quantum dynamics and induced correlations of {Bose} polarons},\ }\href@noop {} {\bibfield  {journal} {\bibinfo  {journal} {New J. Phys.}\ }\textbf {\bibinfo {volume} {22}},\ \bibinfo {pages} {043007} (\bibinfo {year} {2020})}\BibitemShut {NoStop}%
\bibitem [{\citenamefont {Pyzh}\ and\ \citenamefont {Schmelcher}(2020)}]{Pyzh}%
  \BibitemOpen
  \bibfield  {author} {\bibinfo {author} {\bibfnamefont {M.}~\bibnamefont {Pyzh}}\ and\ \bibinfo {author} {\bibfnamefont {P.}~\bibnamefont {Schmelcher}},\ }\bibfield  {title} {\bibinfo {title} {Phase separation of a {Bose-Bose} mixture: Impact of the trap and particle-number imbalance},\ }\href {https://doi.org/10.1103/PhysRevA.102.023305} {\bibfield  {journal} {\bibinfo  {journal} {Phys. Rev. A}\ }\textbf {\bibinfo {volume} {102}},\ \bibinfo {pages} {023305} (\bibinfo {year} {2020})}\BibitemShut {NoStop}%
\bibitem [{\citenamefont {Morera}\ \emph {et~al.}(2021)\citenamefont {Morera}, \citenamefont {Astrakharchik}, \citenamefont {Polls},\ and\ \citenamefont {Juli\'a-D\'{\i}az}}]{dmrg-QD-dimerized-QD-Astrakharchik-PhysRevLett.126.023001}%
  \BibitemOpen
  \bibfield  {author} {\bibinfo {author} {\bibfnamefont {I.}~\bibnamefont {Morera}}, \bibinfo {author} {\bibfnamefont {G.~E.}\ \bibnamefont {Astrakharchik}}, \bibinfo {author} {\bibfnamefont {A.}~\bibnamefont {Polls}},\ and\ \bibinfo {author} {\bibfnamefont {B.}~\bibnamefont {Juli\'a-D\'{\i}az}},\ }\bibfield  {title} {\bibinfo {title} {Universal dimerized quantum droplets in a one-dimensional lattice},\ }\href {https://doi.org/10.1103/PhysRevLett.126.023001} {\bibfield  {journal} {\bibinfo  {journal} {Phys. Rev. Lett.}\ }\textbf {\bibinfo {volume} {126}},\ \bibinfo {pages} {023001} (\bibinfo {year} {2021})}\BibitemShut {NoStop}%
\bibitem [{\citenamefont {Vallès-Muns}\ \emph {et~al.}(2024)\citenamefont {Vallès-Muns}, \citenamefont {Morera}, \citenamefont {Astrakharchik},\ and\ \citenamefont {Juliá-Díaz}}]{dmrg-QD-particle-imbalance-Astrakharchik-10.21468/SciPostPhys.16.3.074}%
  \BibitemOpen
  \bibfield  {author} {\bibinfo {author} {\bibfnamefont {J.}~\bibnamefont {Vallès-Muns}}, \bibinfo {author} {\bibfnamefont {I.}~\bibnamefont {Morera}}, \bibinfo {author} {\bibfnamefont {G.~E.}\ \bibnamefont {Astrakharchik}},\ and\ \bibinfo {author} {\bibfnamefont {B.}~\bibnamefont {Juliá-Díaz}},\ }\bibfield  {title} {\bibinfo {title} {{Quantum droplets with particle imbalance in one-dimensional optical lattices}},\ }\href {https://doi.org/10.21468/SciPostPhys.16.3.074} {\bibfield  {journal} {\bibinfo  {journal} {SciPost Phys.}\ }\textbf {\bibinfo {volume} {16}},\ \bibinfo {pages} {074} (\bibinfo {year} {2024})}\BibitemShut {NoStop}%
\bibitem [{\citenamefont {Tojo}\ \emph {et~al.}(2010)\citenamefont {Tojo}, \citenamefont {Taguchi}, \citenamefont {Masuyama}, \citenamefont {Hayashi}, \citenamefont {Saito},\ and\ \citenamefont {Hirano}}]{Tojo}%
  \BibitemOpen
  \bibfield  {author} {\bibinfo {author} {\bibfnamefont {S.}~\bibnamefont {Tojo}}, \bibinfo {author} {\bibfnamefont {Y.}~\bibnamefont {Taguchi}}, \bibinfo {author} {\bibfnamefont {Y.}~\bibnamefont {Masuyama}}, \bibinfo {author} {\bibfnamefont {T.}~\bibnamefont {Hayashi}}, \bibinfo {author} {\bibfnamefont {H.}~\bibnamefont {Saito}},\ and\ \bibinfo {author} {\bibfnamefont {T.}~\bibnamefont {Hirano}},\ }\bibfield  {title} {\bibinfo {title} {Controlling phase separation of binary {Bose-Einstein} condensates via mixed-spin-channel {Feshbach} resonance},\ }\href {https://doi.org/10.1103/PhysRevA.82.033609} {\bibfield  {journal} {\bibinfo  {journal} {Phys. Rev. A}\ }\textbf {\bibinfo {volume} {82}},\ \bibinfo {pages} {033609} (\bibinfo {year} {2010})}\BibitemShut {NoStop}%
\bibitem [{\citenamefont {Skov}\ \emph {et~al.}(2021)\citenamefont {Skov}, \citenamefont {Skou}, \citenamefont {J\o{}rgensen},\ and\ \citenamefont {Arlt}}]{Skov_fluid}%
  \BibitemOpen
  \bibfield  {author} {\bibinfo {author} {\bibfnamefont {T.~G.}\ \bibnamefont {Skov}}, \bibinfo {author} {\bibfnamefont {M.~G.}\ \bibnamefont {Skou}}, \bibinfo {author} {\bibfnamefont {N.~B.}\ \bibnamefont {J\o{}rgensen}},\ and\ \bibinfo {author} {\bibfnamefont {J.~J.}\ \bibnamefont {Arlt}},\ }\bibfield  {title} {\bibinfo {title} {Observation of a {Lee-Huang-Yang} fluid},\ }\href {https://doi.org/10.1103/PhysRevLett.126.230404} {\bibfield  {journal} {\bibinfo  {journal} {Phys. Rev. Lett.}\ }\textbf {\bibinfo {volume} {126}},\ \bibinfo {pages} {230404} (\bibinfo {year} {2021})}\BibitemShut {NoStop}%
\bibitem [{\citenamefont {Landau}(1933)}]{Landau1933}%
  \BibitemOpen
  \bibfield  {author} {\bibinfo {author} {\bibfnamefont {L.~D.}\ \bibnamefont {Landau}},\ }\bibfield  {title} {\bibinfo {title} {{\"U}ber die bewegung der elektronen in kristallgitter},\ }\href@noop {} {\bibfield  {journal} {\bibinfo  {journal} {Phys. Z. Der Sowjetunion}\ }\textbf {\bibinfo {volume} {3}},\ \bibinfo {pages} {644} (\bibinfo {year} {1933})},\ \bibinfo {note} {also in: Collected Papers of L.D. Landau, 67-68 (1965)}\BibitemShut {NoStop}%
\bibitem [{\citenamefont {Tajima}\ \emph {et~al.}(2021)\citenamefont {Tajima}, \citenamefont {Takahashi}, \citenamefont {Mistakidis}, \citenamefont {Nakano},\ and\ \citenamefont {Iida}}]{Tajima}%
  \BibitemOpen
  \bibfield  {author} {\bibinfo {author} {\bibfnamefont {H.}~\bibnamefont {Tajima}}, \bibinfo {author} {\bibfnamefont {J.}~\bibnamefont {Takahashi}}, \bibinfo {author} {\bibfnamefont {S.~I.}\ \bibnamefont {Mistakidis}}, \bibinfo {author} {\bibfnamefont {E.}~\bibnamefont {Nakano}},\ and\ \bibinfo {author} {\bibfnamefont {K.}~\bibnamefont {Iida}},\ }\bibfield  {title} {\bibinfo {title} {Polaron problems in ultracold atoms: Role of a {Fermi} sea across different spatial dimensions and quantum fluctuations of a {Bose} medium},\ }\href@noop {} {\bibfield  {journal} {\bibinfo  {journal} {Atoms}\ }\textbf {\bibinfo {volume} {9}},\ \bibinfo {pages} {18} (\bibinfo {year} {2021})}\BibitemShut {NoStop}%
\bibitem [{\citenamefont {Knap}\ \emph {et~al.}(2012)\citenamefont {Knap}, \citenamefont {Shashi}, \citenamefont {Nishida}, \citenamefont {Imambekov}, \citenamefont {Abanin},\ and\ \citenamefont {Demler}}]{Knap}%
  \BibitemOpen
  \bibfield  {author} {\bibinfo {author} {\bibfnamefont {M.}~\bibnamefont {Knap}}, \bibinfo {author} {\bibfnamefont {A.}~\bibnamefont {Shashi}}, \bibinfo {author} {\bibfnamefont {Y.}~\bibnamefont {Nishida}}, \bibinfo {author} {\bibfnamefont {A.}~\bibnamefont {Imambekov}}, \bibinfo {author} {\bibfnamefont {D.~A.}\ \bibnamefont {Abanin}},\ and\ \bibinfo {author} {\bibfnamefont {E.}~\bibnamefont {Demler}},\ }\bibfield  {title} {\bibinfo {title} {Time-dependent impurity in ultracold fermions: Orthogonality catastrophe and beyond},\ }\href {https://doi.org/10.1103/PhysRevX.2.041020} {\bibfield  {journal} {\bibinfo  {journal} {Phys. Rev. X}\ }\textbf {\bibinfo {volume} {2}},\ \bibinfo {pages} {041020} (\bibinfo {year} {2012})}\BibitemShut {NoStop}%
\bibitem [{\citenamefont {Cetina}\ \emph {et~al.}(2015)\citenamefont {Cetina}, \citenamefont {Jag}, \citenamefont {Lous}, \citenamefont {Walraven}, \citenamefont {Grimm}, \citenamefont {Christensen},\ and\ \citenamefont {Bruun}}]{Cetina}%
  \BibitemOpen
  \bibfield  {author} {\bibinfo {author} {\bibfnamefont {M.}~\bibnamefont {Cetina}}, \bibinfo {author} {\bibfnamefont {M.}~\bibnamefont {Jag}}, \bibinfo {author} {\bibfnamefont {R.~S.}\ \bibnamefont {Lous}}, \bibinfo {author} {\bibfnamefont {J.~T.~M.}\ \bibnamefont {Walraven}}, \bibinfo {author} {\bibfnamefont {R.}~\bibnamefont {Grimm}}, \bibinfo {author} {\bibfnamefont {R.~S.}\ \bibnamefont {Christensen}},\ and\ \bibinfo {author} {\bibfnamefont {G.~M.}\ \bibnamefont {Bruun}},\ }\bibfield  {title} {\bibinfo {title} {Decoherence of impurities in a {Fermi} sea of ultracold atoms},\ }\href {https://doi.org/10.1103/PhysRevLett.115.135302} {\bibfield  {journal} {\bibinfo  {journal} {Phys. Rev. Lett.}\ }\textbf {\bibinfo {volume} {115}},\ \bibinfo {pages} {135302} (\bibinfo {year} {2015})}\BibitemShut {NoStop}%
\bibitem [{\citenamefont {Goold}\ \emph {et~al.}(2011)\citenamefont {Goold}, \citenamefont {Fogarty}, \citenamefont {Lo~Gullo}, \citenamefont {Paternostro},\ and\ \citenamefont {Busch}}]{Goold}%
  \BibitemOpen
  \bibfield  {author} {\bibinfo {author} {\bibfnamefont {J.}~\bibnamefont {Goold}}, \bibinfo {author} {\bibfnamefont {T.}~\bibnamefont {Fogarty}}, \bibinfo {author} {\bibfnamefont {N.}~\bibnamefont {Lo~Gullo}}, \bibinfo {author} {\bibfnamefont {M.}~\bibnamefont {Paternostro}},\ and\ \bibinfo {author} {\bibfnamefont {T.}~\bibnamefont {Busch}},\ }\bibfield  {title} {\bibinfo {title} {Orthogonality catastrophe as a consequence of qubit embedding in an ultracold {Fermi} gas},\ }\href {https://doi.org/10.1103/PhysRevA.84.063632} {\bibfield  {journal} {\bibinfo  {journal} {Phys. Rev. A}\ }\textbf {\bibinfo {volume} {84}},\ \bibinfo {pages} {063632} (\bibinfo {year} {2011})}\BibitemShut {NoStop}%
\end{thebibliography}%

\end{document}